# Atomic-scale sensing of photoexcitation processes in electronically isolated molecules via atomic force microscopy

Lisanne Sellies[1,2,*], Thomas Buchner[1], Jinhui Guo[3], Sonja Bleher[1], Felix Giselbrecht[1], Clea Ruth[1], Andrea Donarini[4,*], Jascha Repp[1,5,*], Laerte L. Patera[1,3,*]

[1] Institute for Experimental and Applied Physics, University of Regensburg, Regensburg, Germany
[2] IBM Research, Rüschlikon, Switzerland
[3] Department of Physical Chemistry, University of Innsbruck, Innsbruck, Austria
[4] Institute for Theoretical Physics, University of Regensburg, Regensburg, Germany
[5] Halle-Berlin-Regensburg Cluster of Excellence CCE, University of Regensburg, Regensburg, Germany

*Corresponding authors:
lisanne.sellies@ibm.com; andrea.donarini@ur.de; jascha.repp@ur.de; laerte.patera@uibk.ac.at

## Abstract

Atomic-scale insights into light-matter interaction can be obtained using light-assisted scanning probe microscopy techniques. Recently, photoexcited charge carriers have been detected by means of scanning tunnelling microscopy, enabling the study of photo-induced charge transfer with atomic-scale spatial resolution. Here, we propose an approach based on atomic force microscopy (AFM), namely photoexcitation single-charge AFM (PE-AFM), to detect photoexcitation in single molecules adsorbed on non-conductive dielectric films. Synchronizing laser pulses to the oscillation of the tip of an AFM enables the detection of electron tunnelling events that follow photoexcitation. We demonstrate the PE-AFM technique for individual copper phthalocyanine molecules and achieve photoexcitation-driven AFM contrast with ångström spatial resolution. The observed sub-molecular contrast suggests the involvement of a long-lived quadruplet excited state. When combined with the recently developed AFM excited-state spectroscopy including lifetime measurements, PE-AFM enables comprehensive characterization of electronic states involved in photoexcitation and subsequent intersystem crossing, establishing a powerful platform for investigating photophysical processes at the single-molecule level.

## Introduction

Over the last decades, various combinations of scanning tunnelling microscopy (STM) with optical techniques have been developed, including electro[1–3]- and photoluminescence[4,5] and tip-enhanced Raman spectroscopy (TERS)[6–9]. These techniques provide atomic-scale insights into photophysical processes, such as energy transfer[2], coherent intermolecular dipole-dipole coupling[10], photon cascades[11] and photo-induced chemical reactions[12]. Yet, the toolbox to study photophysical processes at the atomic scale misses methods to follow the photoinduced electron-transfer processes, being crucial for energy conversion, for instance in solar cells[13].

Recent pioneering works[14–17] have begun to close this gap by resolving photoinduced electron transfer as an atomically resolved STM tunnelling current. This approach requires the electronic decoupling of molecules from the metal substrate, typically achieved using a 3-4 monolayer (ML) spacer of the wide-bandgap insulator NaCl. The decoupling drastically reduces hybridization with metal states as well as exciton quenching by energy transfer to the metal's conduction electrons[18], often described as electron-hole pair excitation. The increased lifetime of the resulting exciton often remains limited by charge exchange with the substrate, due to the finite junction conductance required for STM. The estimated lifetime resulting from this mechanism is in the mid picosecond and lower nanosecond range on a 3 ML and 4 ML NaCl film, respectively[19,20].

The outstanding significance of photoexcitation processes lies in the resulting charge-separation or its use in photocatalytic processes, for which the scientifically relevant regimes become accessible for fully electrically isolated molecules. The application of AFM on thicker insulating films[21] (for NaCl layers thicker than 14 ML[22,23], in the following referred to as insulating surfaces), that prevent charge exchange with the substrate, opens the door to this regime, while the single-charge sensitivity of AFM[24–26] can be used to detect individual charge-transfer processes on such films[22,27]. In particular, the suppression of substrate tunnelling allows a separation of individual tunnelling processes and enables the temporal control of charge transfer with the tip[28,29]. At the same time, the use of thick films shall avoid the quenching of photoexcitation that is associated with the presence of metallic substrates[18,30].

Here we introduce photoexcitation single-charge AFM as an approach for studying charge-transfer processes at the atomic scale following photoexcitation. Nanosecond laser pulses, tuned to a molecular electronic transition, are used to photoexcite individual molecules, here copper phthalocyanines (CuPc). The photoexcitation drives subsequent tunnelling events between the AFM tip and a molecule underneath, which are detected as an AFM signal. The PE-AFM signals of CuPc point towards the involvement of a long-lived state. Using recently developed methods based on pump-probe voltage-pulse sequences that allow extraction of lifetimes[29] and energies[31] of electronic excited states and a many-body description, we rationalize the pathways involved in the photoexcitation signal, including one via a long-lived quadruplet state. Combining PE-AFM with pump-probe experiments[30,31] thus enables experimental identification of the photoexcitation pathways, as well as the detection of dark states.

## Results

### Near-field enhancement

Existing optically coupled scanning probe techniques heavily rely on strong light field enhancement in the junction[7,32,33]. This enhancement is commonly conceptualized as a gap-plasmon mode confined to the tip–sample junction with sub-nanometer separation between metals[5,6,34,35]. How this concept transfers to insulating surfaces is far from obvious.

For a better understanding, we performed finite-element simulations using COMSOL Multiphysics[36]. Two junction geometries were modelled: a Cu tip positioned 10 Å (i) above a bare Cu substrate, and (ii) above 20 ML NaCl film on Cu substrate. The tip was modelled as a truncated cone terminating in an ellipsoid with an atomistic protrusion approximated by a 5 Å radius sphere (see Fig. 1a, b, Extended Data Fig. 1 and Methods for details). The simulations considered linearly polarized light incident at a polar angle of 5° relative to the surface plane. The light polarization was chosen to be parallel to the sagittal plane containing the optical and the tip axis (p-polarized) in the far field, as the near-field enhancement in a tip–sample junction is much stronger for p-polarized light than s-polarized light (perpendicular to the sagittal plane). The simulations for s-polarized light are presented in Extended Data Fig. 2a-d. The electric field was evaluated in a plane 3 Å above the surface, corresponding roughly to the expected adsorption height of CuPc on NaCl[37]. Whereas p-polarized light produces the largest enhancement in the perpendicular field component[5,6,15,32,33,38], transition dipoles of planar π-conjugated adsorbed molecules lie in the surface plane and couple only to the in-plane electric-field component $E_{\parallel} = \sqrt{E_x^2 + E_y^2}$, as observed in previous experiments[5,10,17,39]. To quantify near-field enhancement, the amplitude of $E_{\parallel}$ was normalized to the incident in-plane field ($E_{0,\parallel}$), arising from the finite polar angle. Fig. 1a and b display the enhancement ratio ($E_{\parallel}/E_{0,\parallel}$) in the *y-z* plane for both geometries. Strong near-field enhancement appears immediately adjacent to the tip in both cases, with peak enhancement ratios reaching 2300 for the Cu tip above Cu substrate (Fig. 1a) and 1200 with the NaCl (20 ML) film (Fig. 1b).

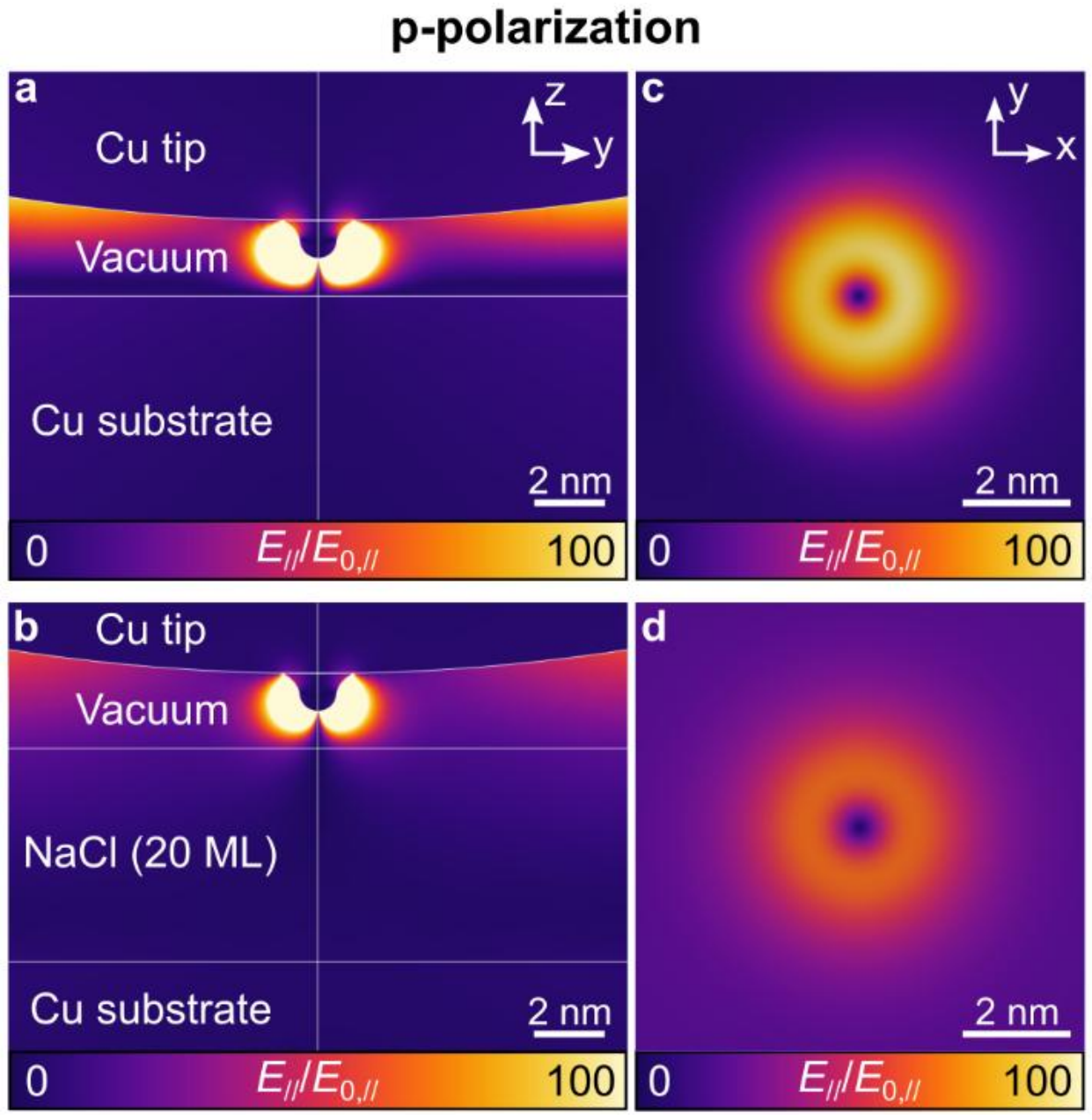

**Fig. 1 | Simulation of the near-field enhancement for p-polarized light**. **a**, **b**, *y-z* cross sections of the enhancement for Cu tip-vacuum-Cu substrate and Cu tip-vacuum-NaCl (20 ML)-Cu substrate, respectively. **c**, **d**, *x-y* cross sections 3 Å above the Cu and NaCl surfaces, respectively. The propagation direction of the light is along the *x* direction. The maximum near-field-enhancement factors are 2300 (**a**), 1200 (**b**), 93 (**c**), and 76 (**d**). In panels **a** and **b**, the signal exceeds the colour-scale maximum and is therefore saturated.

Metals effectively screen electric fields parallel to their surface via free carriers[40]. This is evident in Fig. 1a: screening prevents $E_{\parallel}$ penetration, reducing field amplitude near the vacuum-Cu interface where molecules reside. By contrast, a 20-ML NaCl film lacks free carriers and provides weaker screening. Consequently, $E_{\parallel}$ partially penetrates the dielectric layer with smaller attenuation near the vacuum-dielectric interface. Notably, in the *x-y* plane 3 Å above the substrate, similar enhancement ratios of 93 (bare Cu, Fig. 1c) and 76 (with NaCl, Fig. 1d) are obtained, due to the following competing effects: the Cu-vacuum-Cu system provides larger overall near-field enhancement, but $E_{\parallel}$ is much more strongly attenuated near the surface as compared to the system with the dielectric NaCl layer (see Extended Data Fig. 2e, f). For p-polarization the $E_{\parallel}/E_{0,\parallel}$ amplitude exhibits a ring-shaped (donut) distribution in both systems (Fig. 1c, d) and vanishes directly beneath the tip for symmetry reasons. The resulting signal suppression at the molecular centre[5,10,17,39] therefore provides evidence for direct photoexcitation.

## Concept of photoexcitation single-charge atomic force microscopy

To measure photoexcitation on insulating surfaces using AFM, nanosecond laser pulses are synchronized with the oscillation of the AFM cantilever, such that the tip–sample junction is illuminated only at the zero crossings of the cantilever (see Fig. 2a, b). Laser pulses with a wavelength centred around 640 nm ($E$ = 1.94 eV) drive the optical excitation of the CuPc molecule from $D_0$ to $D_2$[41] (see Methods). This excitation can lead to a single-electron photocurrent between the tip and the molecule, changing the charge state of the molecule (Fig. 2c). The molecule can be neutralized subsequently through electron exchange with the tip (Fig. 2d,e). The tunnelling rate of the discharging process is tuned by the gate voltage $V_G$, which controls the energetic alignment of states with different charges[22,23,31], and can thus be tuned to occur predominantly when the tip is close to the bottom turnaround point (see Fig. 2b). Therefore, the (average) charge state of the molecule differs between the inward and outward motion of the cantilever, as illustrated in Fig. 2c. Different charge states of the molecule result in different forces acting on the cantilever[26], thereby modulating the force at the cantilever's resonance frequency[42], which for the given phase relation can be detected as a damping of the cantilever motion[25,28,43–45]. Since the entire cycle of charging and discharging is initiated by photoexcitation (see Fig. 2c), this provides a scheme to sense photoexcitation. Note that this scheme is analogous to alternate-charging scanning tunnelling microscopy (AC-STM)[28]; however, whereas in AC-STM the tunnelling events are steered by voltage pulses, here they are driven by photoexcitation.

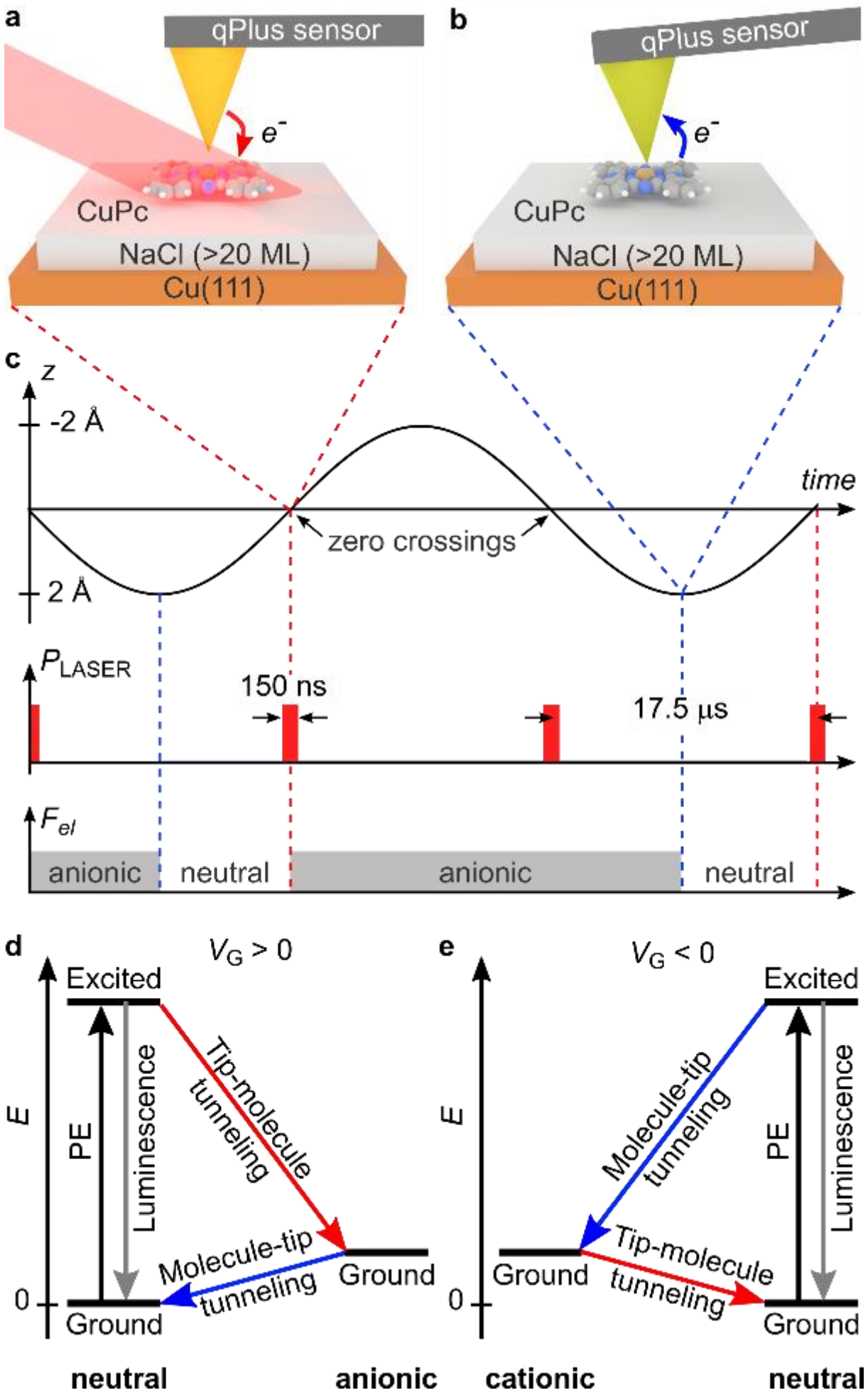

**Fig. 2 | Operating principle of Photoexcited single-charge AFM.** Schematic of the PE-AFM measurement scheme showing nanosecond laser pulses phase-locked to the AFM cantilever oscillation. **a**, Pulses illuminate the tip–sample junction at the cantilever's zero crossings, exciting the CuPc molecule and inducing electron transfer to the molecule (tip) for suitable positive (negative) values of $V_G$. **b**, By setting $|V_G|$ slightly below the charge bistability threshold, molecular neutralization is steered to occur predominantly near the lower turnaround point of the cantilever oscillation. **c**, Charge-state-dependent damping mechanism. The asymmetry in molecular charge state between the inward and outward cantilever motion produces a damping. **d**, **e**, Simplified many-body diagrams illustrating the photoexcitation, and subsequent tunnelling events for positive or negative $V_G$ close to the charge bistability of the neutral and anionic or cationic charge states, respectively. Note that $V_G$ will shift the states by $q\alpha V_G$[23,46], with $q$ being the net charge of the molecule and $\alpha$ the lever arm, accounting for the voltage drop across the different parts of the junction.

## Photoexcitation atomic force microscopy

Fig. 3a illustrates a schematic representation of the experimental setup, where linearly polarized laser light is focused onto the tip–sample junction (see Methods). Fig. 3b shows the photoexcitation-induced AFM damping (detected as a feedback-induced increased driving amplitude $A_{drive}$) signal measured above a CuPc molecule as a function of $V_G$ for different laser polarization angles $\theta$. The signal is maximized for p-polarized light ($\theta = 0°$) and vanishes for s-polarized light ($\theta = 90°$), consistent with the polarization dependence of the near-field enhancement (see Fig. 1 and Extended Data Fig. 2a-d). The strong dependence on the light polarization supports the assignment of the observed damping signals to CuPc photoexcitation, while ruling out alternative effects such as thermal expansion of the tip[47] or direct cantilever excitation by the laser pulses. Regarding the latter, it is important to note that the laser pulses occur at a frequency twice that of the AFM cantilever oscillation (see Fig. 2c), and therefore introduce no spectral component at the cantilever resonance frequency.

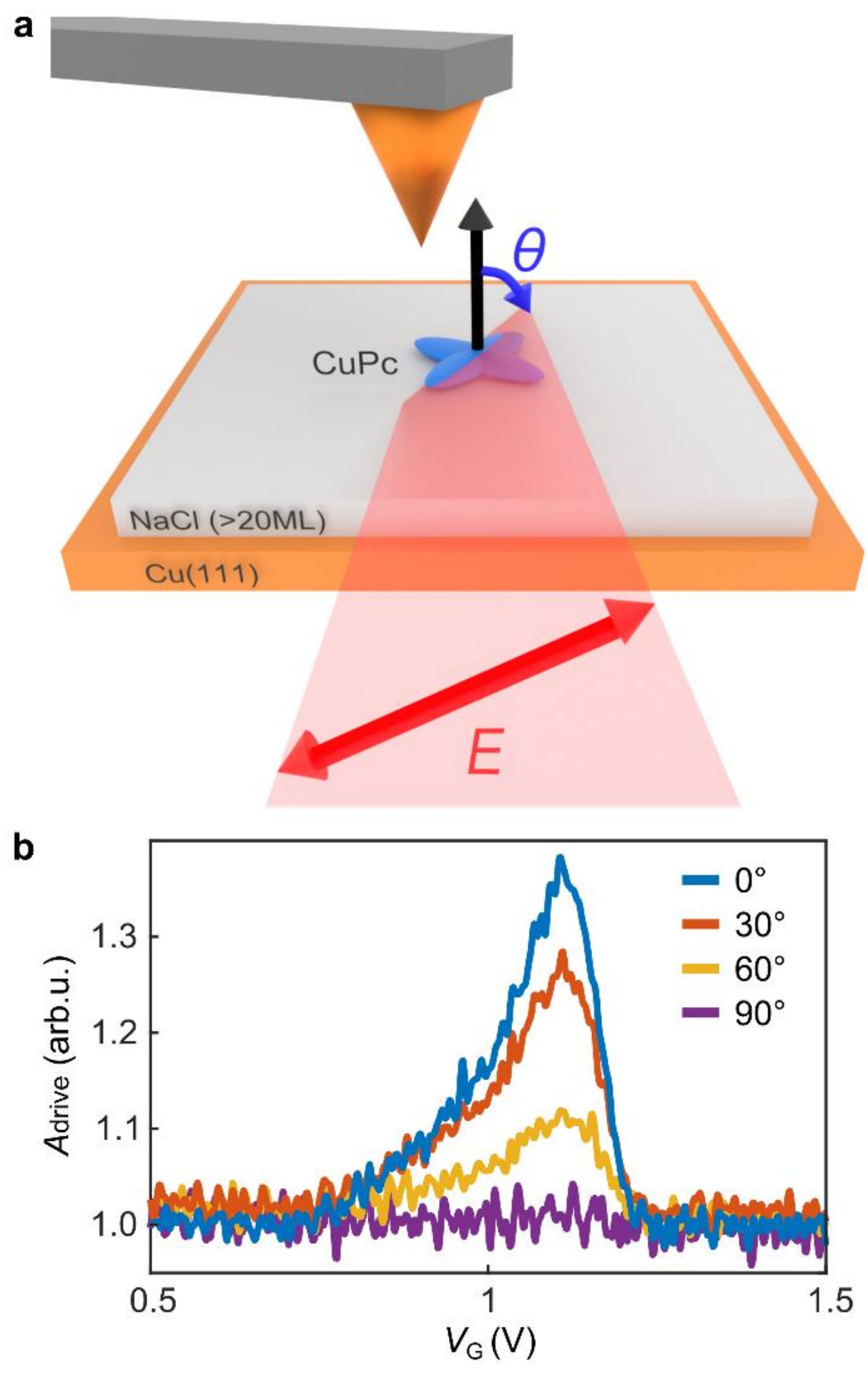

**Fig. 3 | Polarization dependence of the photocurrent signal**. **a**, Schematic indicating the polarization angle $\theta$. **b**, Polarization dependence of the damping signal as a function of $V_G$.

On top of a background, $A_{drive}$ shows pronounced features for positive (Fig. 3b) and negative (Fig. 4a) $V_G$ corresponding to the neutral-to-cationic (Fig. 2e) and the neutral-to-anionic (Fig. 2d) charge-state degeneracy points, respectively. Both features exhibit pronounced asymmetry with a longer tail towards small absolute $V_G$. To understand the characteristic shape, we consider the timing of charging and discharging processes. The charging occurs during and (potentially) after the laser pulse at the cantilever's zero crossing positions, which is followed by a tunnelling event to discharge the molecule. Both of these tunnelling events are affected by $V_G$, but the discharging depends more strongly on $V_G$ because the two relevant states are nearly degenerate (see Fig. 2d, Extended Data Fig. 3). At $V_G$ corresponding to the peak's maxima, the ground states approach degeneracy, which suppresses the discharging rate. As a result, discharging tends to occur, on average, close to the lowest turnaround point, where the tunnelling rate is largest, maximizing damping, and hence $A_{drive}$ (see Fig. 2c). On the long-tail side of the peaks, the discharging rate becomes faster with decreasing $|V_G|$, such that the molecule remains charged for a shorter fraction of the cantilever period and $A_{drive}$ gradually decreases. In direction of increasing $|V_G|$ the discharging rate reduces, and the point will be reached, at which the energy alignment does not allow the molecule to transition into the neutral state anymore. The molecule will simply remain charged over the entire cantilever period, not causing damping of the cantilever oscillation, which explains the relatively sharp drop on this side of the peaks.

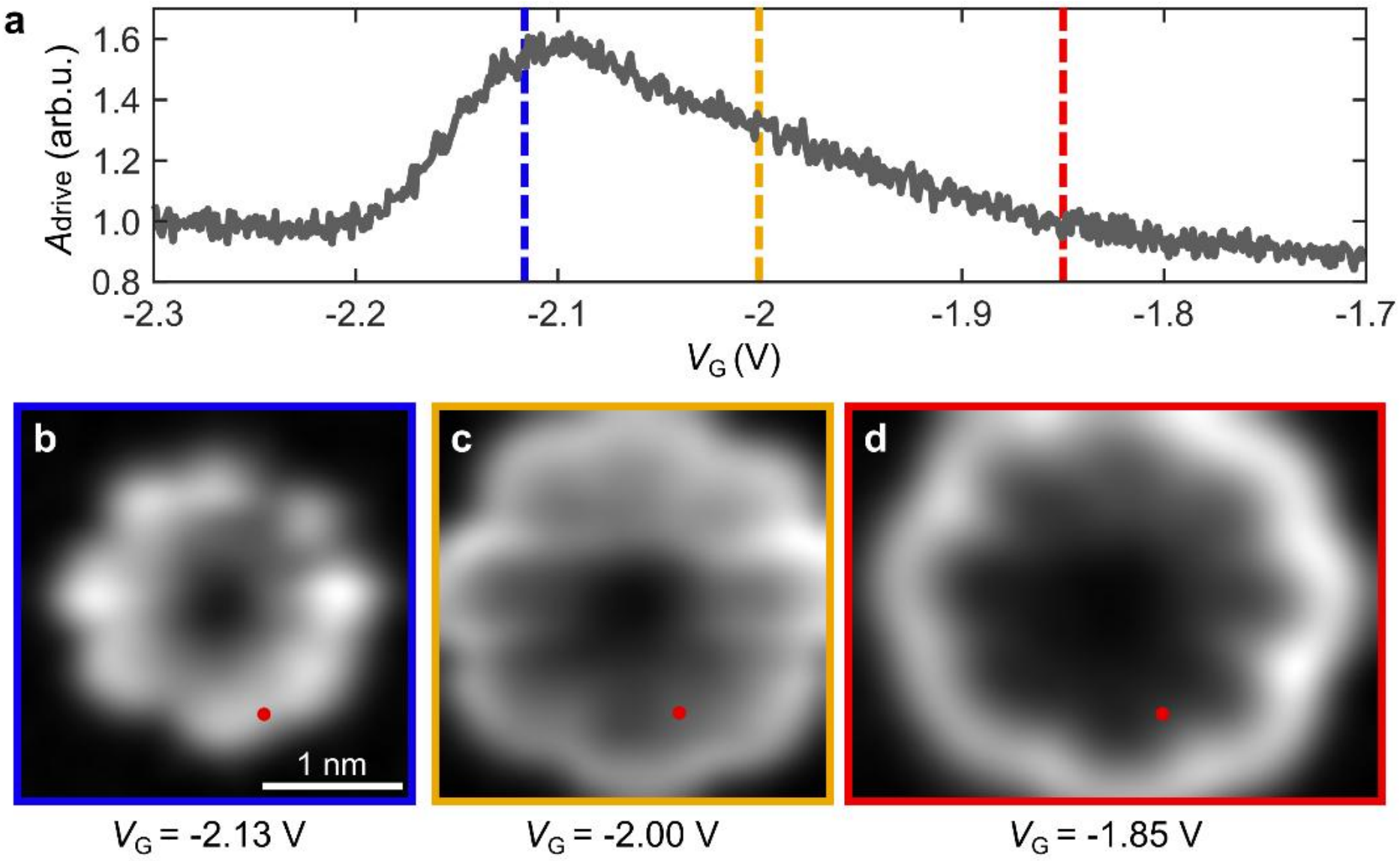

**Fig. 4 | Gate voltage control of photoexcitation detection in CuPc. a**, $V_G$-dependent $A_{drive}$ (indicating damping) for the neutral-to-cationic transition showing an asymmetric lineshape due to the interplay between fixed photoexcitation timing and $V_G$-tunable discharging dynamics. **b**–**d**, Constant-height PE-AFM images ($A_{drive}$) at $V_G$ = –2.13 V (**b**), –2.00 V (**c**), and –1.85 V (**d**). The red dot indicates the position where the spectrum shown in a is taken.

We mapped the PE-AFM signal, that is, $A_{drive}$ spatially in constant-height mode for different $V_G$ (see Fig. 4b-d). At $V_G$ = -1.85 V a ring-like contrast is observed at the periphery of the molecule (Fig. 4d). At higher absolute gate voltages ($V_G$ = -2.00 V, Fig. 4c), additional signal appears around the centre of the molecule, and at even higher gate voltages ($V_G$ = -2.13 V, Fig. 4b) only signal is visible around the centre, corresponding to the region where the frontier orbital densities are the highest. This dependence on $V_G$ can be explained similarly as for the point spectra in Fig. 3a and 4a, when considering that the tunnelling rate into the molecule is large directly above the molecule and exponentially decreases when moving laterally away: For low $|V_G|$, the discharging rate is too fast around the centre of the molecule, while for high $|V_G|$ the discharging rate is too low at the edge.

Notably, the contrast of the PE-AFM image in Fig. 4b differs significantly from both the highest occupied (HOMO) and lowest unoccupied molecular orbital (LUMO) densities measured in CuPc molecules by AC-STM (see Extended Data Fig. 4)[28,48]. While the signal intensity is observed at similar locations, the contrast, although preserving sub-molecular features, is smeared and lacks a well-defined nodal plane structure. This is in contrast with photoexcitation measurements performed with STM for a free-base phthalocyanine ($H_2Pc$)[15] and for other molecules[16,17], where a contrast resembling the orbital density of the excited state could be imaged. A similar flattening of orbital-density contrast was recently observed by Kaiser *et al.* in STM experiments in which the tunnelling current through a molecular resonance was saturated[20]. By analogy, we hypothesize that the contrast observed in the PE-AFM images likewise arises from a signal saturation due to a long-lived state – a hypothesis substantiated further below.

Moreover, the PE-AFM images shown in Fig. 4 exhibits a central depression above the molecule, in line with the spatial profile of the tip-enhanced in-plane field (see Fig. 1c and d), similarly as observed in STM-based photoluminescence studies[5,17]. This observation rules out hot-electron generation in the tip followed by tunnelling from/to the molecule[17] as an alternative mechanism for the PE-AFM contrast, thereby further pointing to photoexcitation as its origin. Additional evidence to rule out hot-electron generation is the absence of excess $A_{\mathrm{drive}}$ once the molecule is positively or negatively charged (see Extended Data Fig. 5a, b): hot-electron excitation in the tip should work largely independent of the molecule's charge state[17]. In contrast, resonant photoexcitation of the molecule strongly depends on its charge state[3]. Furthermore, control experiments on pentacene yielded no detectable photoexcitation signal, consistent with the laser energy (1.94 eV) being lower than the optical gap (2.26 eV)[49], lending further support for the interpretation as a photoexcitation signal in the case of CuPc.

## Electronic structure and photoexcitation pathways for CuPc

To clarify the photoexcitation processes and the photocharge relaxation, we characterized the relevant ground and excited states of CuPc by means of excited-state spectroscopy (see Methods for details)[31]. The resulting simplified many-body diagram is shown in Fig. 5a, as explained in detail in the Methods. In its electronic ground state CuPc exhibits an open-shell configuration characterized by a metal-centred singly occupied molecular orbital (SOMO), resulting in a doublet ($D_0$). The low-lying excited states of the neutral molecule are either doublet ($D_1$, $D_2$) or quartet states ($Q_1$), retaining a single electron in the SOMO (see Methods section). The lifetime of the $Q_1$ state has been measured using the pump-probe voltage pulse sequence reported by Peng et al.[29] (see Fig. 5b and Methods) as 11 ± 3 μs, attributed to the $Q_1$ to $D_0$ spin-forbidden transition.

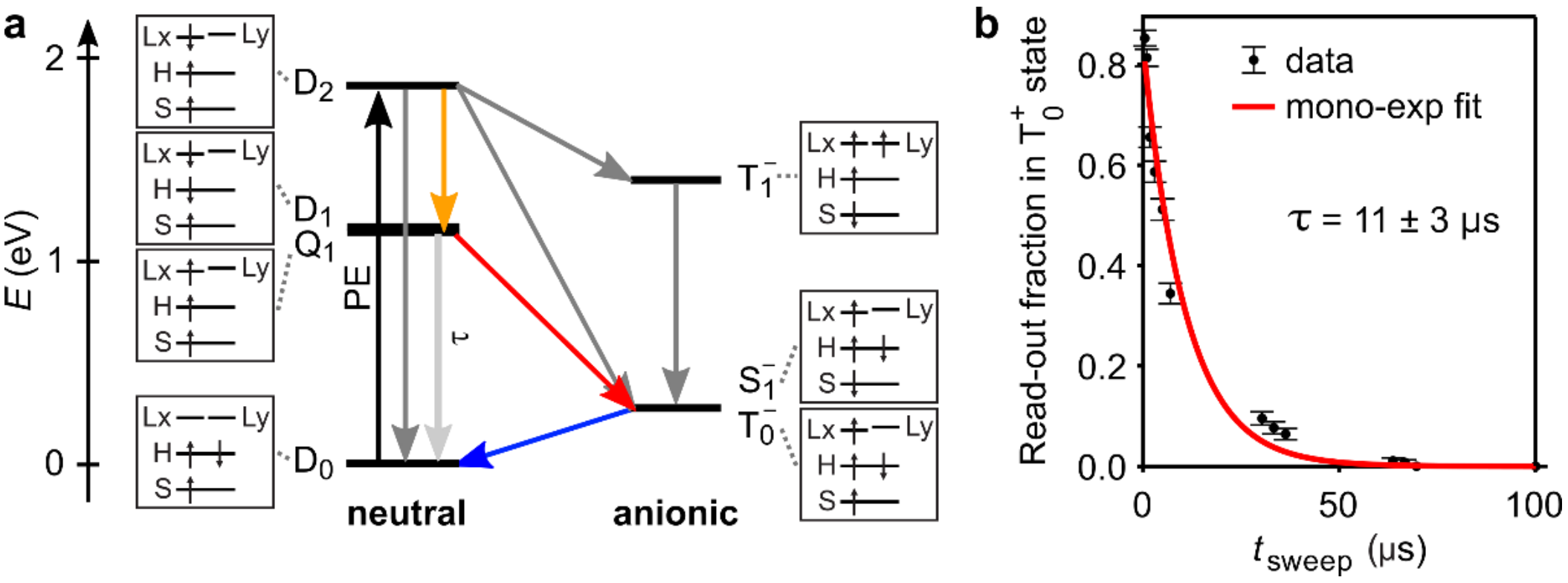

**Fig. 5 | Energy level diagram and $Q_1$ lifetime determination of CuPc**. **a**, Many-body diagram showing relevant electronic states determined by excited-state spectroscopy (see Methods). The insets depict the electronic configurations, that is, the occupation of the SOMO (bottom), HOMO (middle) and two (degenerate) LUMOs (top). One representative electronic arrangement is shown for each configuration, and for multiconfiguration states only one component is shown (see Methods). **b**, $Q_1$ state lifetime $\tau$ = 11 ± 3 μs measured via pump-probe spectroscopy. The fraction read out in the $T_0^+$ state, reflecting the $Q_1$ population, is plotted versus the duration of the sweep pulse.

The experimental observation of a long-lived $Q_1$ state provides a consistent picture for the observed photoexcitation signal: during photoexcitation, the molecule can transition into the long-lived $Q_1$ state via intersystem crossing[50,51] (see Fig. 5a, and Methods for more details), which could lead to a tunnelling event significantly delayed with respect to the short duration of the laser pulse. To experimentally confirm that such delayed tunnelling events can occur, we increased the AFM amplitude in PE-AFM experiments from 2 to 9 Å (see Extended Data

Fig. 5c). The tip-height was adjusted such that the closest turnaround point of the tip remained at the same distance to the sample, while the zero crossing moved away from the sample by up to 7 Å. If the tunnelling only occurred quasi-instantaneous with photoexcitation at the zero crossings, the tunnelling probability would be reduced by roughly seven orders of magnitude. In contrast, the PE-AFM signal was only moderately reduced with increasing amplitude, in line with a long-lived dark state and delayed tunnelling. For this to work, the state's lifetime must be at least an appreciable fraction of the cantilever period, in agreement with the lifetime measurements (Fig. 5b).

Because of its long lifetime, it is reasonable that the pathways via $Q_1$ dominate the photoexcitation contrast, as discussed in more detail in the Methods. This provides a natural explanation of the saturation of the charging probability per cycle, leading to the observed flatting of the orbital density contrast (see Methods). Put simply, charging following photoexcitation remains possible for such a prolonged time, that it can proceed almost independent of the exact tip position. Instead, the discharging tunnelling event can become the limiting process, such that it may dominate the weak contrast on the largely saturated signal. We confirmed this by performing experiments on a CuPc molecule in which the degeneracy of the two LUMOs is lifted by the local molecular environment, allowing us to distinguish between tunnelling events involving the LUMO and HOMO, see Extended Data Fig. 6 and Methods.

## Discussion

Our results demonstrate that single-electron photocurrents can be detected by AFM with submolecular spatial resolution for molecules adsorbed on insulating surfaces. Finite-element simulations (Fig. 1) and experimental data (Fig. 3) reveal substantial near-field enhancement at the tip–sample junction, enabling efficient photoexcitation even when molecules are electronically isolated from the underlying metal substrate by 20 ML of NaCl.

The absence of clear nodal plane structure observed for the PE-AFM image of CuPc (Fig. 4b) provides clear indication of the involvement of a long-lived state ($Q_1$, $\tau$ = 11 ± 3 μs) in the tunnelling processes governing the PE-AFM contrast. Whereas photoexcitation signals had been measured for PtPc with STM-based photoluminescence and electroluminescence, the triplet-state lifetime was found to lie between 2.9 ps and 1 ns, being dramatically shorter than the one measured for PtPc in DMF solutions (≈ 1.55 μs), attributed to charge exchange with the underlying metal[52]. In contrast, our observation exemplifies that PE-AFM is applicable to a sample environment, which largely preserves the long lifetimes of states following photoexcitation. PE-AFM has thus the potential to probe single-molecule photoexcitation processes involving long-lived excited states close to their intrinsic transition rates.

Both STM[15–17] and AFM photocurrent detection rely on a cascade of photophysical processes that collectively determine the spatial contrast in images. However, the nature of the tunnelling processes involved varies. Photocurrent detection via STM involves both molecule-tip and molecule-substrate tunnelling, with the latter governed primarily by intrinsic molecule-substrate coupling and thus remaining largely insensitive to applied bias or tip position. Instead, in PE-AFM, all tunnelling events occur between the tip and the molecule, with the advantage that they can be controlled by the applied gate voltage. The timing of these tunnelling processes with respect to the cantilever motion is crucial for the observed signal, and all tunnelling processes depend on the tip position. This level of control makes PE-AFM particularly suited to investigate and disentangle competing photophysical mechanisms at sub-molecular length scales.

Future developments of PE-AFM will focus on extending the method toward different charge states, easily accessible by the applied gate voltage on insulating surfaces[22,27], enabling systematic investigations of how different charge configurations influence photoexcitation pathways and relaxation dynamics. Combined with a pulsed laser source with tunable wavelengths, PE-AFM could be applied to a broader class of molecular systems. More complex pulse patterns, optionally combined with gate-voltage pulses open the door to pump–probe–type experiments, to follow excited-state population dynamics, resolve competing relaxation channels, and disentangle sequential tunnelling processes with unprecedented detail. Beyond single-molecule studies, PE-AFM can be employed to explore charge[22,43] and energy transfer[2] within and between molecular assemblies, providing insight into intermolecular coupling and collective photophysical behaviour.

# Methods

## Simulations

A three-dimensional (3D) model was constructed in the Wave Optics Module of COMSOL Multiphysics 6.4,[36] which employs the finite element method (FEM), to solve frequency-domain Maxwell´s equations for the electric field:

$$\nabla \times \mu_r^{-1}(\nabla \times \vec{E}) - k_0^2 \left(\epsilon_r - \frac{j\sigma}{\omega\epsilon_0}\right)\vec{E} = 0$$

In the simulation, a linearly polarized plane wave with the wavelength of $\lambda = 640\,nm$ propagated at a 5° angle with respect to the x-axis.

The concept of constructing a tip with an atomic-scale protrusion was adopted from Yang *et al.*[5], who proposed that a small silver (Ag) atomic cluster acting as a protrusion on top of a Ag tip (~ 50 nm in radius) significantly enhances the nanocavity plasmonic field, extending the field confinement down to the sub-atomic scale. In our model, a Cu tip was used, being represented by a truncated cone with a bottom ellipsoid. The cone had the bottom radius $r_{\text{cone}} = 50\,nm$, a height $h_{\text{cone}} = 100\,nm$ and a semi-angle $\frac{\theta}{2} = 10°$. The vertical distance between the bottom of the cone and the bottom of the ellipsoid was defined as $h' = 40\,nm$. To ensure a smooth joint between the cone and the ellipsoid, the following equations were imposed to determine the ellipsoid radius ($r_{\text{ellip}}$) and height ($h_{\text{ellip}}$)[53]:

$$\frac{{r_{\text{cone}}}^2}{{r_{\text{ellip}}}^2} + \frac{(h_{\text{ellip}} - h')^2}{{h_{\text{ellip}}}^2} = 1$$

$$\frac{\theta}{2} = \frac{\pi}{2} - \tan^{-1}\left[\frac{{h_{\text{ellip}}}^2 r_{\text{cone}}}{{r_{\text{ellip}}}^2 (h_{\text{ellip}} - h')}\right]$$

The atomistic protrusion of the Cu cluster was approximated as a small sphere with a radius of 0.5 nm attached to the tip apex[5]. Geometric continuity between the protrusion and the cone was enforced by revolving an interpolation curve defined from the centre plane of the sphere to the tip apex.

A Cu substrate of a radius and a height of 250 nm and an identical Cu substrate covered with 20 ML NaCl layer with a radius of 50 nm were built in the model. The gap between the tip and the Cu or 20ML-NaCl/Cu substrate was 1 nm. The tip and substrate were surrounded by a spherical simulation domain of 700 nm radius. A perfectly matched layer (PML) with a thickness of 400 nm terminated the outer boundary of the domain. Convergence tests validated the chosen sizes of the simulation sphere and the Cu cylinder substrate. The complex dielectric constants for Cu and for NaCl were adopted from the literature.[54,55]

The Swept meshes were chosen for the PML domain, while the Free Tetrahedral meshes were employed in all other domains. The number of elements in the Swept mesh was validated via a convergence study to ensure effective absorption of the electromagnetic waves without spurious reflection. The Free Tetrahedral mesh sizes were limited to 64 nm ($\lambda/10$) to adequately resolve the electromagnetic wavelength for accurate field calculations. In the atomistic protrusion and the gap region between the tip and substrate, much finer meshes of within 0.1 nm were applied.

To quantify the electric field enhancement, the values were averaged over a disc area with a radius of 1 nm, located at a height of 3 Å above the substrate, corresponding to the size and adsorption height of a CuPc molecule.

**STM/AFM setup**

Experiments were carried out with two home-built combined STM/AFM microscopes equipped each with a qPlus sensor[56] and a conductive Pt-Ir tip in ultrahigh vacuum. The PE-AFM experiments were performed on one of the microscopes (resonance frequency, $f_0$ = 28.5 kHz; spring constant, $k \approx 1.8$ kNm$^{-1}$; quality factor, $Q \approx 1.0 \times 10^5$) and the excited-state spectroscopy measurements on the other similar microscope (resonance frequency, $f_0$ = 30.0 kHz; spring constant, $k \approx 1.8$ kNm$^{-1}$; quality factor, $Q \approx 1.9 \times 10^4$). The microscopes were operated under ultrahigh vacuum (base pressure, $P < 10^{-10}$ mbar) at $T \approx 8$ K (in the absence of laser irradiation) in frequency-modulation mode, in which the frequency shift $\Delta f$ of the cantilever resonance is measured. The cantilever amplitude was 1 Å (2 Å peak-to-peak) for the excited-state spectroscopy and AC-STM[28] measurements and 2 Å for the photoexcitation measurements, except if stated differently. PE-AFM and AC-STM images were taken in constant-height mode, at a reduced tip height as indicated by the negative $\Delta z$ values (tip-height change with respect to the AFM setpoint).

**Sample preparation**

The surface of a Cu(111) single crystal was cleaned by repeated cycles of neon ion sputtering and thermal annealing to 550 °C. After the last sputtering, the crystal was annealed to 500 °C. Afterwards, approximately 25 ML of NaCl were evaporated from a crucible onto half of the sample, while the other half was covered by a mask to ensure clean areas of metal for tip preparation. The temperature of the sample during NaCl evaporation was kept at 100 °C. Copper phthalocyanine (CuPc) molecules were deposited via flash evaporation onto the cold sample ($T \approx 8$ K) inside the scan head. CuPc molecules adsorb in different geometries. Here, those with the Cu ion positioned above a $Na^+$ site of the NaCl(001) surface, known to adopt a stable adsorption geometry,[37,41] were investigated, except if stated differently.

**Laser setup**

The laser setup is schematically illustrated in Extended Data Fig. 7a. A nanosecond pulsed laser (NPL; NPL64C, Thorlabs) with a centre wavelength of 640 nm was used. The NPL was triggered by the AFM controller via an arbitrary waveform generator (AWG) to synchronize the laser pulses with the AFM cantilever oscillation. The AWG was used to adjust the phase and to initiate two laser pulses per cantilever oscillation. The NPL beam is sent through two polarizers, of which the first is rotatable. Since the output of the NPL is linearly polarized, the pair of polarizers can be used to adjust the power coupled to the SPM junction. After the polarizers, a $\lambda/2$ wave plate was used to rotate the polarization of the laser. Except where stated otherwise, the polarization was adjusted to maximize the resulting PE-AFM signal[57] (see Fig. 2). Following the wave plate, the laser beam was widened and recollimated using a set of two lenses. The resulting beam diameter of approximately 15 mm leads to a higher numerical aperture when focusing the laser to the tip, which will in turn lead to a smaller focus spot. Using a 90:10 beam splitter, 10% of the power was diverted to a photodiode to monitor the timing of the laser pulses.

The beam was guided to the STM chamber by mirrors, of which the last one is rigidly mounted to the UHV chamber. A lens was mounted to a motorized 3-axis stage, which allows focusing the beam onto the tip–sample junction, as has been demonstrated by Schröder et al.[58] The beam's polar incident angle is about 5° with respect to the surface plane. The laser beam has been reflected from the metallic sample and focused on another photodiode at the other side of the UHV chamber, which was used for laser alignment and monitoring the stability of the overall laser power. Extended Data Fig. 7c shows a recorded photodiode signal as a function of the lens position scanned perpendicular to the beam path (see Extended Data Fig. 7b), which was used to focus the laser onto the tip–sample junction.

The laser pulses had a duration of ≈ 150 ns. Two laser pulses were applied per cantilever period (≈ 35 μs), to avoid excitation of the cantilever[28], resulting in an on-off ratio of ≈ 1 : 117. We observed a temperature increase from 8 K to 9.2 K for typically used average laser power of 8.5 mW. Note that 8.5 mW is the power of the illumination into the system, a large fraction of it leaves the cryostat after being reflected at the sample, such that the power absorbed by the microscope head is lower. An even much smaller fraction of the total power is reaching the junction, because of the low numerical aperture in the experiment. After turning on the laser, a three-hour thermalization period was allowed before the experiments were performed.

**Excited-state spectroscopy of CuPc**

The excited-state spectroscopy measurements were performed and analyzed as explained in ref. 31, with the adjustments as detailed below for CuPc. The resulting spectra are shown in Extended Data Fig. 8.

The excited-state spectroscopy data was fitted analogously to the data for pentacene and PTCDA described in ref. 31. As a start, transitions as illustrated in Extended Data Fig. 9d, e were included, based on the expected low-lying electronic states for CuPc, as discussed in detail in the Methods section und *Electronic structure of CuPc*. Spin-allowed transitions within one charge state were assumed to be instantaneous (e.g. $D_1 \rightarrow D_0$) compared to the tunnelling rates. Intersystem crossing was not explicitly taken into account. Note, however, because of the similar energy of $D_1$ and $Q_1$, intersystem crossing from $D_1$ to $Q_1$ would be captured by our fitting as a reduced rate of tunnelling into $D_1$ and increased rate of tunnelling into $Q_1$. Three fit parameters per transition were included: the threshold voltage, at which the transition opens or closes, the rate and the width of the transition.

The fitting procedure described in ref. 31 assumes that all the population in the long-lived $Q_1$ state transfers to either $T_0^+$ or $T_0^-$ during the read-out phase, depending on whether the read-out occurs at the $T_0^+$-$D_0$ or $D_0$-$T_0^-$ degeneracy, respectively. This assumption holds if the tunnelling rates of $Q_1 \rightarrow T_0^+$ and $Q_1 \rightarrow T_0^-$ are fast compared to the decay rate of $Q_1$. In CuPc, however, the $Q_1$ decay rate is sufficiently fast that it cannot be neglected relative to the $Q_1 \rightarrow T_0^-$ rate. In contrast, it remains negligible compared to the faster $Q_1 \rightarrow T_0^+$ rate (the tunnelling barrier is smaller for $Q_1 \rightarrow T_0^+$ compared to $Q_1 \rightarrow T_0^-$ since it is a LUMO tunnelling process instead of a HOMO tunnelling process). We take this into account into the fitting by introducing a fixed parameter $f_{\mathrm{read-out}}$ such that the read-out fraction in the $T_0^-$ state, $T^-_{0,\mathrm{read-out}}$, is given by

$$T^-_{0,\mathrm{read-out}} = T_0^- + f_{\mathrm{read-out}}(Q_1 + f_{+\rightarrow -}T_0^+).$$

If $f_{\mathrm{read-out}} = 1$, $T^-_{0,\mathrm{read-out}}$ is a sum of the populations in $T_0^-$ and $Q_1$ as well as a fraction $f_{-\rightarrow +}$ of the population in $T_0^+$ that gets transferred into $T_0^-$ via $Q_1$ (given by the ratio of $k_2/k_1$, see Extended Data Fig. 9d, e). $1 - f_{\mathrm{read-out}}$ reflects the fraction of $Q_1$ that decays into $D_0$ instead of transferring to $T_0^-$ at the beginning of the read-out period.

The parameter $f_{\mathrm{read-out}}$ can be directly obtained from the excited-state spectra (Extended Data Fig. 8). It is given by the ratio of the plateau around -0.6 V for read-out at the $D_0$-$T_0^-$ hysteresis (Extended Data Fig. 8d) to that for the $T_0^+$-$D_0$ hysteresis (Extended Data Fig. 8c) (for a given sweep time). Importantly, the cantilever oscillation (at 1 Å amplitude, that is, 2 Å peak-to-peak, to optimise the signal-to-noise ratio for charge-state detection[59]) modulates the tunnelling rates. To minimize effects on the resulting spectra, the sweep voltage pulses were synchronized with the cantilever oscillation period, such that the entire pulse occurs around the smallest tip-sample distance. For the longest of the four sweep times, the cantilever has time to move away from the closest turnaround point, reducing the $Q_1 \rightarrow T_0^+$ rate in the subsequent read out. Accordingly, a smaller value of $f_{\mathrm{read-out}}$ was used for the longest sweep time. $f_{\mathrm{read-out}}$ for the longest sweep time was scaled by the ratio of the plateau around -2.3 V (Extended Data Fig. 8b, d) for the longest sweep time to that for the shorter sweep times.

Likewise, $f_{\mathrm{read-out}}$ had to be adapted for initialization in $T_0^+$ compared to initialization in the neutral states, since the longer initialization sequence for the preparation in the neutral state led to a different timing of the start of the read-out period with respect to the cantilever's oscillation.

We assume a lever arm $\alpha$ of 0.70 ± 0.05, corresponding to the values derived in ref. 31 for pentacene on 20 ML of NaCl. Analogous to ref. 31, in the fitting equations (see Supplementary Information of ref. 31), we set the work function of the tip and NaCl covered Cu(111) surface to 4.2 eV, the tunnelling distance to 9 Å and an upper limit to the fitted widths of 0.23 V. In conventional scanning tunnelling spectroscopy, the phonon-broadened states are observed as Gaussian peaks in the differential conductance[60], corresponding to an onset of the tunnelling rate with voltage as an error function[23], in which the full width at half maximum (FWHM) can be obtained by multiplying the fitting parameter for the width with $2\sqrt{\ln(2)}$. The upper limit to the width is required since in some cases the fit could compensate for a too large width of a transition by using a larger rate. Since the widths of all the transitions that do not show this behaviour are similar and around 0.17 V for CuPc (FWHM 0.28 V), we set a slightly larger upper limit for all widths of 0.23 V (FWHM 0.38 V).

Next to the transitions shown in Extended Data Fig. 9d, e, we included two additional transitions to improve the fit. We note that these additional transitions are not the focus of this work and were included solely to improve the fitting of the main transitions. This mainly affects the fit above $V_{\mathrm{sweep}}$ = 1.5 V for the shortest sweep time. The best fit was obtained with one additional transition from $T_0^+$ into a long-lived neutral state, as well as a transition from $Q_1$ to a higher lying anionic state. The choice of the quenching via one higher-lying anionic state for one additional transition improves the fit significantly for the data measured with neutral initialization. The difference between a transition from $T_0^+$ into a long- or a short-lived neutral state is minimal, with a slightly better fit for a transition from $T_0^+$ into a long-lived state. Since the fit did not significantly improve by adding more than two additional transitions, we chose to only add two additional transitions. Note that there are other states that are accessible in the experiment, such as the $D_2$ state, but their effect on the measured data is too small to detect them.

The fitted voltages at which the transitions occur are displayed for every transition in Extended Data Table 1, with error bars given by the uncertainty provided by the fitting algorithm. Analogous to ref. 31, we set a lower limit on the uncertainties of the fitting parameters to ± 0.05 V to account for uncertainty derived from fit robustness checks (by varying the parameters). In addition, we tested the influence of the additional transitions on the fitting voltages of the other transitions, by changing the number and nature of transitions taken into account. Further, we tested the influence of fitting the data in a narrower voltage range. Most transitions were not affected within their uncertainty margins, except for $T_0^+ \rightarrow Q_1$, $Q_1 \rightarrow T_1^-$ and $D_0 \rightarrow T_0^-$, for which we increased the error bars to 0.10 based on this analysis. Note that even when $f_{\mathrm{read-out}}$ was varied to an extent that clearly degraded the fit quality (e.g., by 0.02), the fitted parameters remained unchanged within their error margins. The uncertainty margins on the gaps and reorganization energies (95% coincidence interval) were derived from the uncertainties of the transition voltages and the uncertainty on the lever arm $\alpha$ = 0.70 ± 0.05. Next to the dataset presented in Extended Data Fig. 8, we included measurements performed above one of the lobes instead of the centre of the same molecule, as well as measurements performed above the centre of a second molecule in Extended Data Table 1. In addition, we performed measurements on a CuPc molecule on NaCl(>20 ML)/Au(111) and on a CuPc molecule adsorbed with its metal centre located above a $Cl^-$ site of NaCl(>20 ML)/Cu(111). In those two cases, we obtained the same values within uncertainty margins for the derived gaps. Note that the $Q_1$-$D_0$ gap can be derived in three different ways from the fitted transition voltages in Extended Data Table 1; the obtained values match within error margins. The value obtained from the difference between the $T_0^+ \rightarrow Q_1$ and $T_0^+ \rightarrow D_0$ transitions is listed in Extended Data Table 1. The uncertainty margins on the averaged energy values (95% coincidence interval) were determined from the standard deviation on the three individual values, taking the uncertainty on the lever arm $\alpha$ into account.

### Lifetime measurement

The lifetime measurement of CuPc on NaCl(> 20 ML)/Au(111) was performed with the voltage pulse sequence introduced in ref. 29. This voltage-pulse sequence is similar as the one for the excited-state spectroscopy measurements, consisting of a set, a sweep and a read-out phase, during which specific gate voltages were applied. In case of CuPc, the set pulse initializes the molecule in the $T_0^+$ state ($V_{\text{set}}$ = -2.489 V, $t_{\text{set}}$ = 333.5 μs), while at the beginning of the sweep pulse ($V_{\text{sweep}}$ = 0.511 V) this population gets transferred into the $D_0$, $Q_1$ and $D_1$ states. The population of $D_1$ is assumed to decay instantaneously on the timescale of our pulses, partly into $D_0$ and partly via intersystem crossing into $Q_1$. During the sweep pulse, $Q_1$ can decay into $D_0$. At the beginning of the read-out phase ($V_{\text{read-out}}$ = -1.489 V), the remaining population in the $Q_1$ state is mapped onto the $T_0^+$ state, while the population in the $D_0$ state remains in the $D_0$ state. A read-out time of 125 ms was used to allow extracting the charge state from the measured AFM frequency shift. The pump-probe voltage pulse sequence was repeated 640 times for each sweep time to determine the read-out fraction (the normalized population during the read-out phase) in the $T_0^+$ state, and thereby the $Q_1$ population. The error bars were derived from the standard deviation for a binominal distribution[29].

### Electronic structure of CuPc: calculation, experimental results and literature

The electronic structure of the cationic, neutral and anionic CuPc is largely determined by the occupation and mutual interaction of its four frontier orbitals: i.e. the SOMO, the HOMO and the doubly degenerate LUMOs. The orbital designations HOMO, SOMO, and LUMO always refer to the corresponding orbitals of the neutral molecule. The SOMO is a molecular orbital strongly localized on the metallic centre, the HOMO resides, due to its symmetry, only on the ligand while the LUMOs are distributed over the entire molecule. In order to find the electronic structure, we write, for this reduced set of orbitals, an interacting Hamiltonian which includes all symmetry allowed Coulomb integrals along the lines presented in ref. 61. For the neutral molecule, three electrons are dynamically occupying the frontier orbitals, while the other valence electrons are frozen in the lower-lying molecular orbitals. In particular, due to its localized character, the SOMO only accommodates one electron, despite its single particle energy being lower than the one of the doubly occupied HOMO. Such disagreement with a simple Aufbau principle stems from the large difference in charging energy between the two orbitals. The latter privileges the full occupation of the HOMO against the one of the SOMO despite the energy loss of promoting an electron from the SOMO to the HOMO. The single occupation of the SOMO also characterizes the ground state of the cationic and the anionic molecule, as in ref. 23.

With respect to the model presented in ref. 61, we explicitly include here the crystal field renormalization added to the single-particle energy of the frontier orbitals. The latter reflects the imbalance of the Mulliken charge associated to the electrons in the fully occupied frozen molecular orbitals, which generates a strongly positive metallic centre. For simplicity we take only two values of the crystal field renormalization

$$\Delta E_{\text{HOMO}} = \Delta E_{\text{LUMO}} = 1.83 \text{ eV}, \ \Delta E_{\text{SOMO}} = -3.7 \text{ eV}.$$

The normalization for the HOMO and the LUMO has been determined by fitting the anionic and cationic resonances to experimental measurements on CuPc on thin insulating films, as reported in ref. 63. Because of the positive charge on the metal we improved the model in ref. 63 by choosing a negative value for the SOMO. The latter further stabilizes the single occupation of the SOMO with respect to not only the addition but also to the removal of an electron. With respect to the results in ref. 63, the many-body energy eigenstates with an empty SOMO are now shifted to energies outside the relevant energy range of the presented PE-AFM findings. Excited states with double occupation of the SOMO would only appear in the relevant energy range for an even stronger negative renormalization of the SOMO single particle energy (at $\Delta E_{\mathrm{SOMO}} \approx -6\ \mathrm{eV}$). Of these excited states, the ones with lowest energies are an excited doublet neutral state with the third electron occupying the HOMO and a singlet anionic state with double occupation of SOMO and HOMO. Including those states, though, would not influence the results presented in our work: the mentioned excited neutral doublet cannot be photoexcited from the neutral ground state due to the vanishing transition dipole moment between SOMO and HOMO. Moreover, transitioning to this state from the anionic ground states requires the simultaneous tunnelling of two electrons, which can be neglected under our experimental conditions. Similarly, the anionic state with double occupation of SOMO and HOMO cannot be accessed from the $D_2$ or $Q_1$ state by a single-electron tunnelling process.

Following a recent study on the impact of the substrate on the Jahn-Teller effect characterizing the $D_{4h}$ symmetric CuPc[62], we characterize the many-body eigenstates in terms of the adiabatic potential surfaces (APES) calculated as a function of a generalized coordinate Q combining the Jahn-Teller active modes of the molecule and the ones of the substrate.

In Extended Data Fig. 9a-c, we show an overview of the APES for the cationic, neutral and anionic molecule. The local minima indicated by the dots correspond to the low energy many-body eigenstates for the molecule, with a colour-coded spin multiplicity. For example, the ground state of the neutral CuPc molecule has an open-shell configuration with a metal-centred SOMO, resulting in a doublet $D_0$. As mentioned above, the low-lying excited states of the neutral molecule are either doublet or quartet states, retaining a single electron in the SOMO. The first excited states are the doubly orbital degenerate quartets $Q_{1,x}$ and $Q_{1,y}$, closely followed by the doubly degenerate doublets $D_{1,x}$ and $D_{1,y}$ and, eventually, by the doublets $D_{2,x}$ and $D_{2,y}$. The index x or y indicates an imbalance in the occupation of the LUMO state with an x or a y character. Such an imbalance is favored by a corresponding deformation of the molecule along the antisymmetric Jahn-Teller active modes as indicated by the position of the minima along the horizontal axis.

The energy differences between the electronic ground state and the $Q_1$ and $D_1$ states have been experimentally determined to be 1.13 ± 0.12 eV and 1.09 ± 0.13 eV, respectively. Notably, in excited-state spectroscopy no feature was observed that could be assigned to the transition into the $D_2$ state (based on the expected energetic alignment[41]). Consistent with our PE-AFM results and other experiments on CuPc[50,51], we assume that an appreciable fraction of $D_2$ population will decay quasi-instantaneously into $D_1$ and from there via intersystem crossing into $Q_1$. Note that this could drastically reduce the visibility of the excited-state spectroscopy signal of $T_0^+ \rightarrow D_2$: Upon accessing $D_2$, part of the population is expected to transfer into $Q_1$, which could result in similar read-out signals as for accessing $Q_1$ and $D_1$ directly at lower voltages before opening the $T_0^+ \rightarrow D_2$ transition. The $D_2$ energy we inferred from the STML measurements by Doležal et al. as 1.86 eV with respect to $D_0$.[41]

Because of the Jahn-Teller effect,[62] the degeneracy is lifted upon occupation of the LUMOs with one electron, either by injecting one electron into the molecule,[37] or by photoexcitation from the ground state. As explained above, the low-lying anionic states retain a single electron in the SOMO and are, therefore, either singlet, triplet or quintet states. We experimentally derived the energy of an excited negatively charged state, assigned to the $T_1^-$ and/or $Q_1^-$ state, as 1.21 ± 0.09 eV, with respect to the negatively charged ground state $T_0^-$. Because of a moderate overlap of the SOMO and the ligand-centred HOMO and LUMO, the singlet state $S_1^-$ is only slightly higher in energy than $T_0^-$ by approximately 0.02 eV[63]. Our excited-state spectroscopy results are in good agreement with literature results that cover individual aspects of the spectrum[50,41]. For instance, our values are close to those reported by Vincett et al. for CuPc molecules in 1-chloronaphthalene solution: 1.13 eV and 1.15 eV for the $D_0 \rightarrow Q_1$ and $D_0 \rightarrow D_1$ transitions, respectively[50]. The small discrepancies we attribute to differences in the molecular environment.

**Multiconfiguration states, absorption and radiative decay**

The many-body Hamiltonian for the frontier orbitals used for the description of CuPc contains exchange terms, as well as pair hopping and split pair hopping terms[61]. It is thus expected for the energy eigenstates to have a multiconfigurational character. Exemplarily we shall consider the low energy doublets of the neutral molecule, i.e. the ground state $D_0$ as well as the second and third excited states $D_{1,x}$ and $D_{2,x}$. Their multiconfigurational representations, restricted to the largest contributions, read

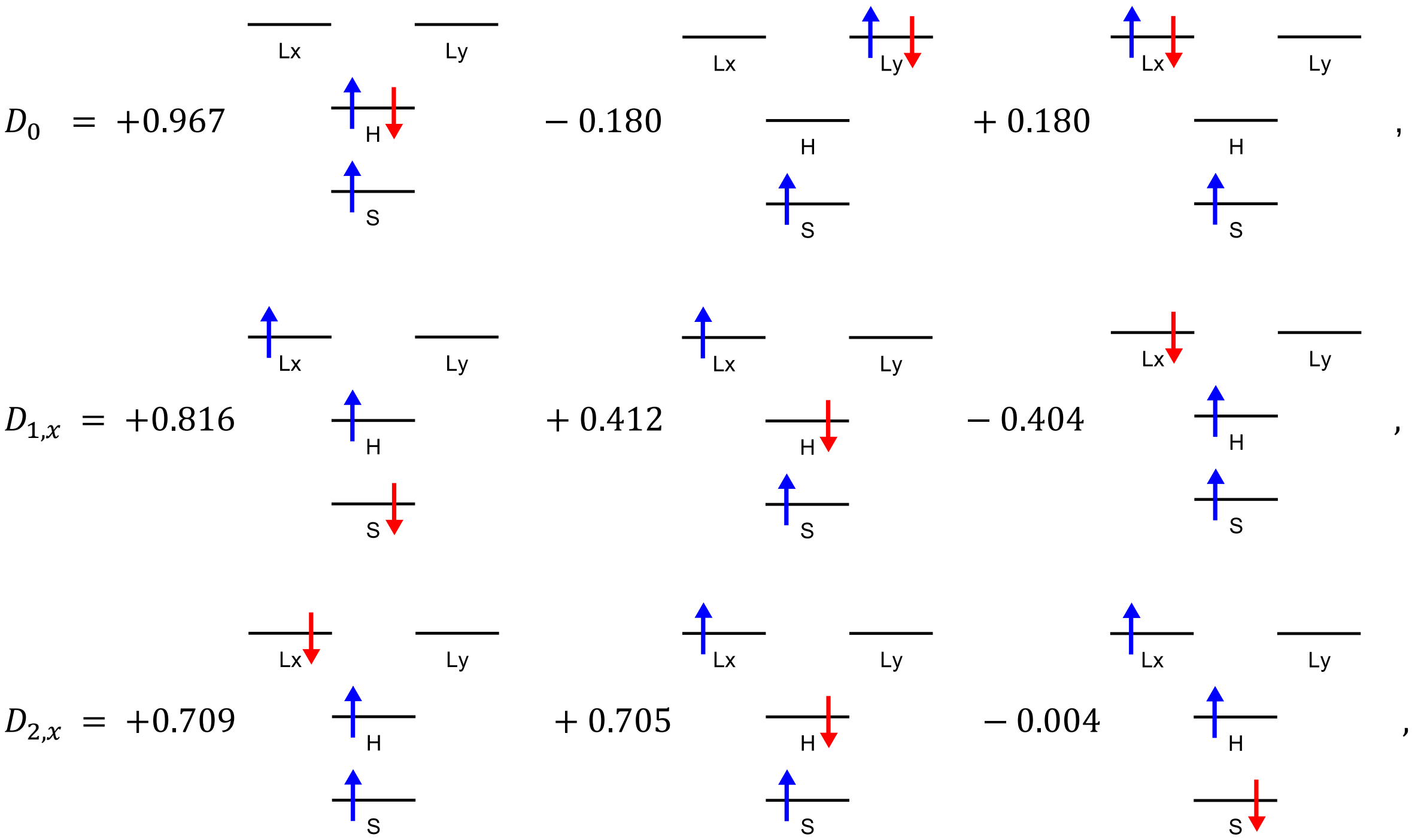

where, for simplicity, only the spin-up states have been represented.

The multiconfigurational composition of the many-body states is crucial for the understanding of the transition rates induced by coupling to the electromagnetic radiation. Due to the spatial symmetry of the frontier molecular orbitals, an electric field polarized along the y direction induces transitions between the HOMO and the LUMOx, while conserving the spin of the electron, thus coupling $D_0$ to both $D_{1,x}$ and $D_{2,x}$. The intensity of the coupling, though, depends crucially on the amplitude and sign of the multiconfigurational components. For example, the first component of the ground state $D_0$ couples to the second and third component of $D_{1,x}$ and to the first and second component of $D_{2,x}$, respectively. For what concerns the transition $D_0 \rightarrow D_{1,x}$ the coupling amplitudes are rather low and moreover interfere destructively due to their opposite sign. In contrast, the transition $D_0$-$D_{2,x}$ is characterized by two components with rather high amplitudes, moreover interfering constructively. We can thus conclude that both in the absorption as well as in the radiative decay, the transition $D_0$-$D_{2,x}$ dominates the picture.

### Pathways for photoexcitation signal generation and saturated contrast

Figure 5a shows the potential pathways during photoexcitation of CuPc. The photoexcitation brings the molecule into the optically accessible $D_2$ state during the 150 ns laser pulse. Aside from a decay back to $D_0$, either directly or via $D_1$, the many-body diagram indicates three other pathways from the $D_2$ state to the ground state at the applied gate voltage. All of the three include a net charge change and may therefore contribute to the resulting damping signal. Two pathways include a direct transition from $D_2$ into the anionic state, either via $T_0^-$ or $T_1^-$. In addition, $D_2$ may intersystem cross into $Q_1$, likely via $D_1$.

The transitions from $D_{2,x}$ or $D_{2,y}$ to $Q_{1,x}$ or $Q_{1,y}$ are likely happening via internal conversion from $D_2$ to $D_1$ followed by an intersystem crossing between $D_1$ and $Q_1$. The radiative decay between $D_2$ and $D_1$ is in fact completely suppressed, as both states have the same occupation of the frontier orbitals, although combining their multi-configurational components with different amplitudes. A relatively high internal conversion rate of $D_2$ to $D_1$ is expected from the crossing between the APES of $D_{2.x}$ and the one of $D_{1,y}$ at small deformations and small excess energies (see Extended Data Fig. 9b). The same holds true for $D_{2.y} \rightarrow D_{1,x}$. The intersystem crossing rate for the $D_1 \rightarrow Q_1$ transition is also expected to be large due to the small energy splitting (of the order of 20 meV) between the involved states, as compared to the one of about 760 meV (moreover underestimated by the present theoretical model) between the long-lived $Q_1$ and $D_0$.

After intersystem crossing from $D_2$ into $Q_1$, subsequent tunnelling into $T_0^-$ may occur during the 11 µs lifetime of $Q_1$. In contrast, the other tunnelling pathways ($D_2 \rightarrow T_0^-$ and $D_2 \rightarrow T_1^-$) are only open during the 150-ns-lasting laser pulse. Hence, the time during which the pathway via $Q_1$ is open is a factor 100 longer than for the other transitions, meaning that even a small occupation of $Q_1$ can lead to a significant contribution to the image contrast.

Once the molecule is trapped in $Q_1$, the decay from $Q_1$ to the ground state competes with the $Q_1 \rightarrow T_0^-$ transition. At the tip height used in the PE-AFM measurements, the rates of this tunnelling transition were on the order of 100 ns. Even if the tunnelling rate was reduced by one order of magnitude because of the lateral tip position (i.e. above a nodal plane of the involved orbital), the $Q_1 \rightarrow T_0^-$ transition is still fast compared to the lifetime of $Q_1$. Hence, the variation of the tunnelling rate due to lateral tip position has only a small effect on the signal. This explains a uniform contrast, as we observed.

With these considerations, we anticipate that the saturated photoexcitation contrast can be explained by the dominating effect of the $D_2 \rightarrow D_1 \rightarrow Q_1 \rightarrow T_0^- \rightarrow D_0$ pathway, leading to saturation due to the long lifetime of $Q_1$. The contrast for negative gate voltages can be explained analogously.

### Influence of the Jahn-Teller effect on photoexcitation

The degeneracy of the LUMO levels together with the Jahn-Teller effect increases complexity, but at the same time widens the accessible experimental parameter space. The two transition dipoles of the two degenerate $D_0$-$D_2$ photoexcitation transitions are both in the molecular plane, but perpendicular to one another. Just as the in-plane component vanishes altogether for the tip centred above the molecule (see Figs. 1c, d and 4), its coupling to one of the two transition dipoles will vanish along the entire vertical symmetry plane running along two opposing iso-indole units. However, the molecule can be photoexcited to either of the LUMOs depending on the tip position, such that the images still carry the four-fold rotational symmetry. The four-fold-symmetric incoherent superposition of the two LUMO-densities is somewhat similar to the HOMO density, such that the influence of tunnelling transitions involving the HOMO and the LUMO are difficult to distinguish from the images.

To further investigate the observed PE-AFM contrast and the influence of the Jahn-Teller effect, we performed experiments on a CuPc molecule, in which the degeneracy of the two LUMOs is lifted by the local molecular environment. The AC-STM image in Extended Data Fig. 6a demonstrates that the degeneracy lifting is already present in the neutral state, as evidenced by the emergence of two-fold symmetry, in contrast to the four-fold symmetry characteristic of the $0 \rightarrow 1^-$ transition (Extended Data Fig. 4a).[28] PE-AFM images acquired at positive and negative gate voltage are shown in Extended Data Fig. 6c and d. The image at positive gate voltage appears to exhibit a reduced symmetry, indicating that a tunnelling process involving the LUMO (short: LUMO tunnelling) is dominating the image contrast. In contrast, the map for negative gate voltage does not show this reduced symmetry, hinting that the contrast is mainly determined by HOMO tunnelling. Hence, in both cases, the discharging process is dominating the contrast, as is illustrated in the many-body diagrams in Extended Data Fig. 6e and f. It is always the tip-position-dependent rate-limiting process of an entire cycle of transitions that dominates the contrast. Apparently, in the present case, the discharging process is the rate-limiting one. This is in line with the cycles presented in Extended Data Fig. 6e and f involving the long-lived $Q_1$ state: the $D_2 \rightarrow Q_1$ transitions do not involve tunnelling and will therefore barely depend on the tip position and are expected to occur with a high rate (see previous section), the neutral → charged tunnelling transition is expected to occur much faster than the reverse, because its larger energy difference exceeds the relaxation energy of this transition. Interestingly, the degeneracy of the two LUMOs is already lifted by the local molecular environment for this molecule, such that one of the two previously equivalent photoexcitation routes becomes energetically favored. However, we observe a four-fold symmetry at negative gate voltages (Extended Data Fig. 6c). If only one photoexcitation channel remains (to one of the $D_2$ states), a nodal plane should be observed perpendicular to the corresponding transition dipole moment, as explained above and reported for the photoexcitation of pentacene[17]. In other words, we seem to photoexcite to both $D_2$ states, yielding a similar photoexcitation AFM intensity, despite that we expect the transitions to differ in energy. Thus, the two-fold symmetry of the PE-AFM map in Extended Data Fig. 6d can be assigned to the tunnelling process following photoexcitation and not to the photoexcitation process itself.

## Data availability

The data that support the plots within this paper and other findings of this study are available from the publication server of the University of Regensburg.

## Acknowledgements

We thank Tzu-Chao Hung, Leo Gross, Katharina Kaiser, and Guillaume Schull for discussions and Philipp Scheuerer, Andreas Rank, Christoph Rohrer, Philipp Thurau, Hugh Lohan for support. Financial support from the European Research Council (ERC) under the European Union's Horizon 2020 research and innovation programme (L.L.P. and J.G. under grant agreement No. 101039746; L.S., S.B., F.G. and J.R. under grant agreement No. 951519, "MolDAM") and from the Swiss National Science Foundation (L.S. under Swiss Postdoctoral Fellowships grant No. 233895, "ISOTOPE") is gratefully acknowledged.

## Author's contributions

L.L.P. conceived the original experimental idea. L.S., T.B., J.R., and L.L.P. designed the experimental details of the study. L.S., T.B. and L.L.P. performed the PE-AFM experiments and L.S., S.B., F.G. and C.R. performed the excited-state spectroscopy experiments. L.S., T.B., S.B., F.G., C.R., J.R., and L.L.P. analyzed the experimental data. J.G. and L.L.P performed and analyzed the electromagnetic field-enhancement simulations. A.D. developed the many-body theoretical description and derived and rationalized the selection rules. L.S., A.D., J.R., and L.L.P. integrated the overall interpretation and wrote the manuscript. All authors discussed the results and contributed to the manuscript.

## Competing interests

The authors declare no competing interests.

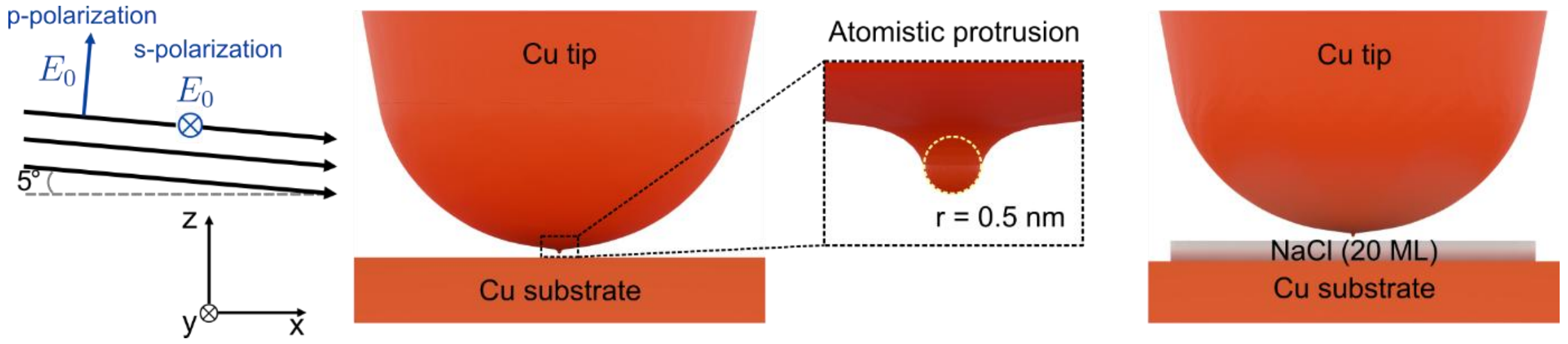

**Extended Data Fig. 1 | Geometries used for the simulation of the near-field enhancement.** (left) Cu tip-vacuum-Cu substrate and (right) Cu tip-vacuum-NaCl(20 ML)-Cu substrate. Inset: zoom-in on the atomistic protrusion at the tip apex.

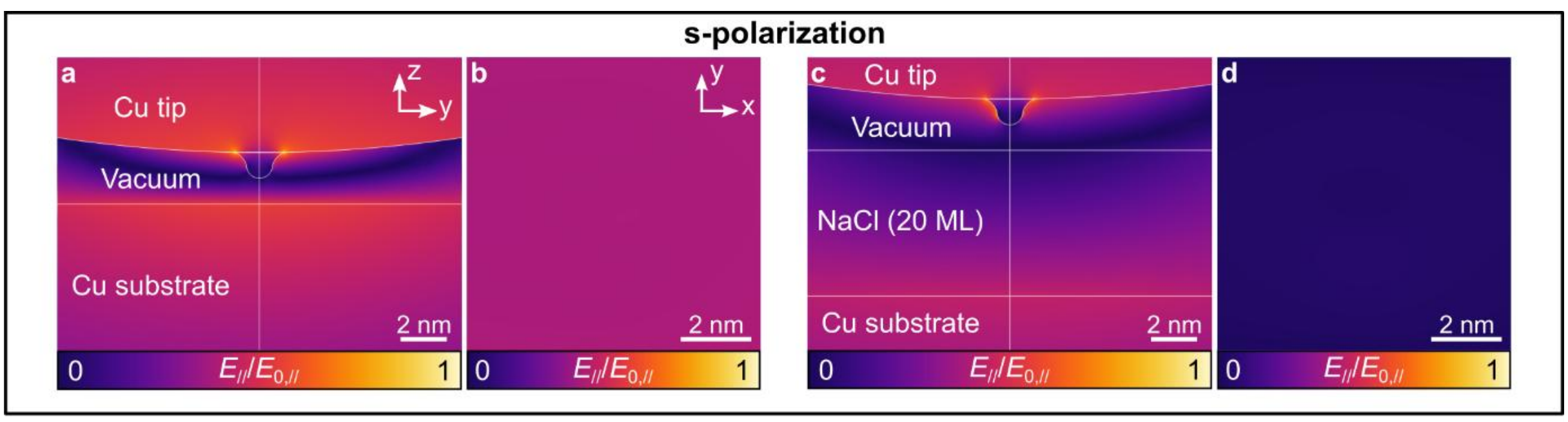

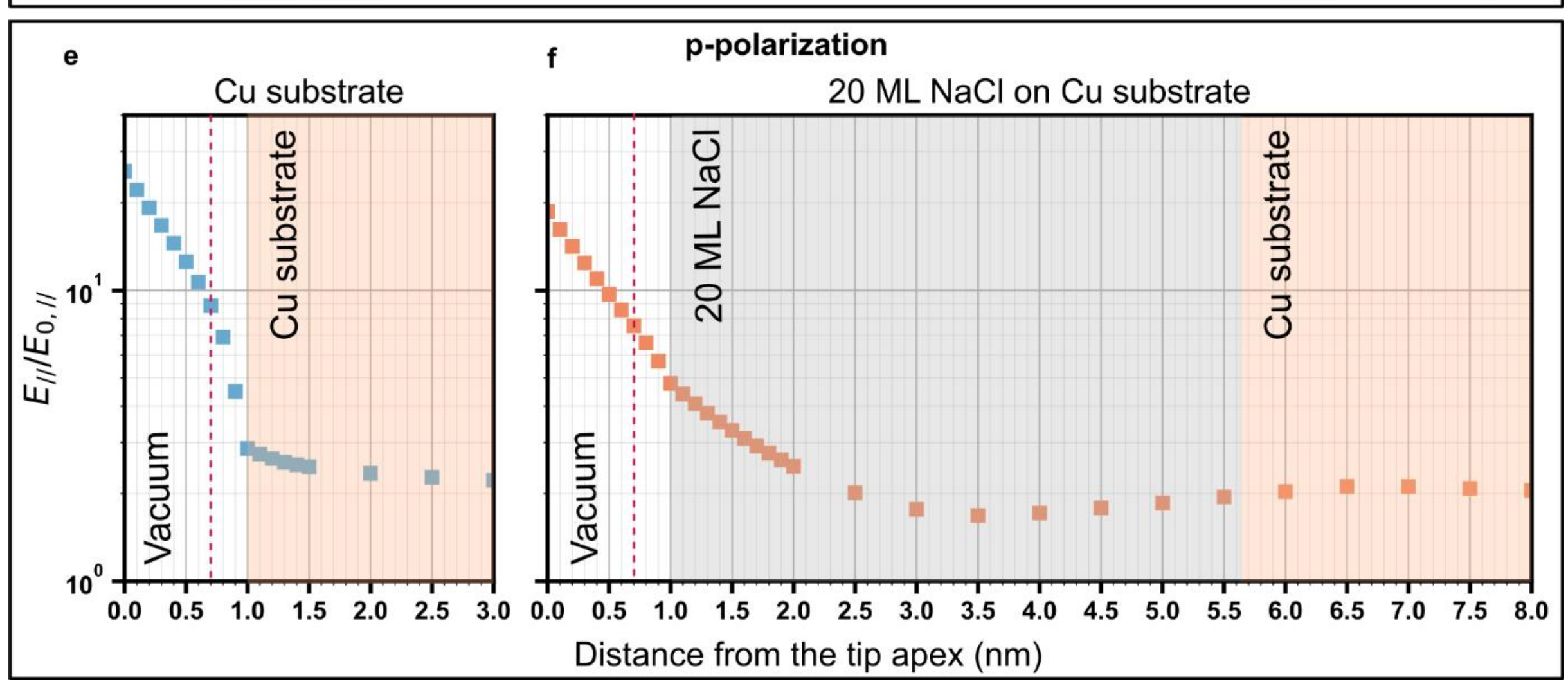

**Extended Data Fig. 2 | Simulation of the near-field enhancement**. **a**, **b**, Plots for Cu tip-vacuum-Cu substrate in the *y*-*z* plane (**a**), and *x*-*y* plane 3 Å above the Cu surface (**b**), respectively, for s-polarized light. **c**, **d**, Plots for Cu tip-vacuum-NaCl (20 ML)-Cu substrate in the *y*-*z* plane (**c**), and *x*-*y* plane 3 Å above the Cu surface (**d**), respectively, for s-polarized light. The propagation direction of the light is along the *x* direction. In panels **a** and **b**, the signal exceeds the colour-scale maximum and is therefore saturated. **e, f** In-plane enhancement ratio ($E_{\parallel}/E_{0,\parallel}$) for p-polarized light as a function of distance to the tip apex for Cu tip-vacuum-Cu substrate (**e**) and Cu tip-vacuum-NaCl (20 ML)-Cu substrate (**f**) respectively. Because for p-polarization the in-plane enhancement vanishes directly beneath the tip (see Fig. 1c, d), the $E_{\parallel}/E_{0,\parallel}$ values were obtained by averaging over a circle of radius 1 nm centred on the tip axis.

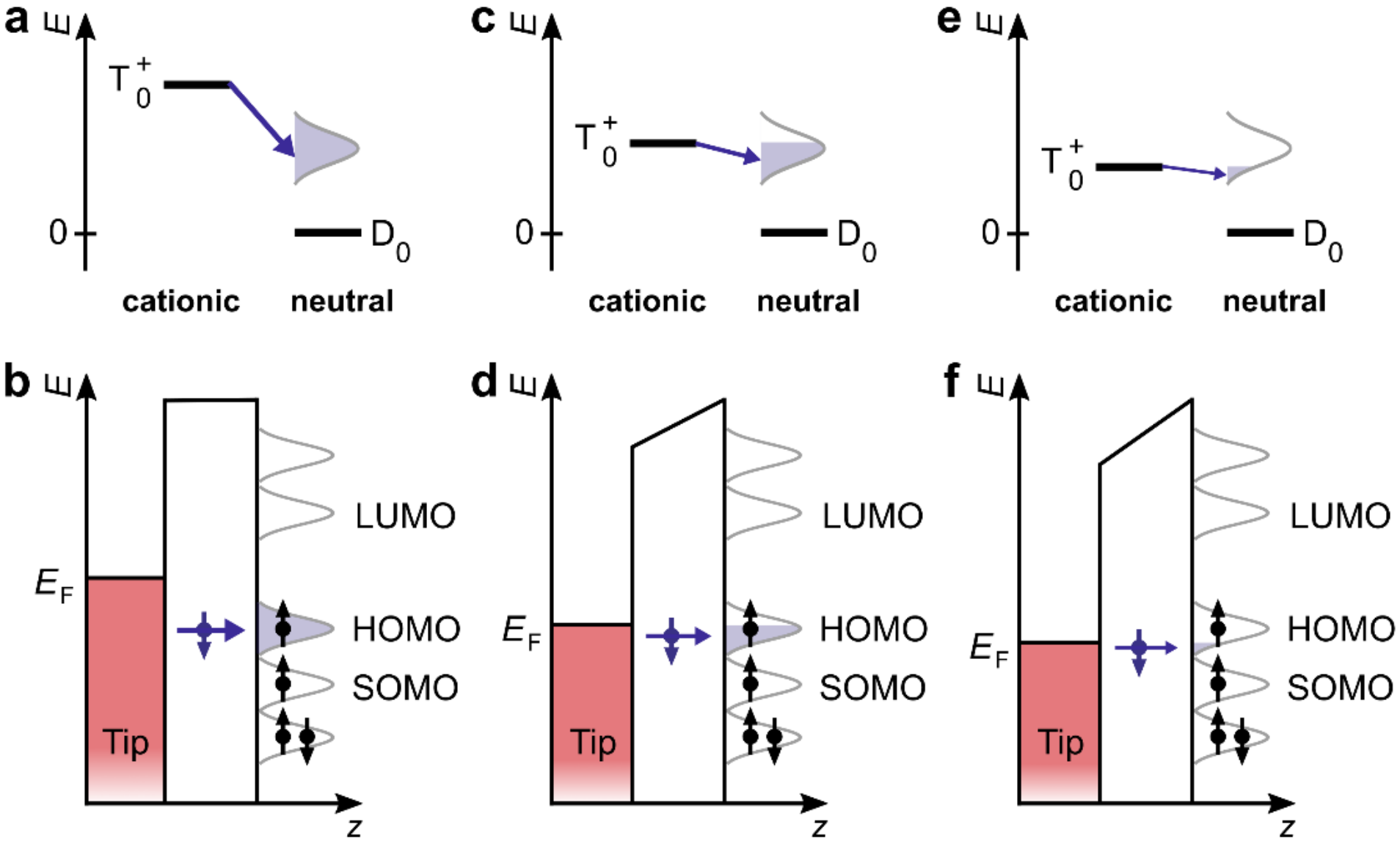

**Extended Data Fig. 3 | Effect of gate voltage on the tunnelling rate for transitions between states close to degeneracy. a**, **c**, **e**, Many-body diagrams for the tunnelling transition between two states differing in their net charge, here exemplarily $T_0^+$ and $D_0$ states of CuPc, at three different gate voltages. $V_G$ becomes increasingly more negative from **a** to **e**, shifting the cationic state downwards ($V_G$ shifts the states by $q\alpha V_G$[23,46], with $q$ being the net charge of the molecule and $\alpha$ the lever arm, because part of $V_G$ drops across the NaCl film). Because of the strong coupling between the net charges in the adsorbed (CuPc) molecule and the phonons in the ionic film (NaCl), there is a vanishing Franck-Condon overlap between the vibrational ground states of the two charge transitions (black lines). Consequently, for a gate voltage, for which the states are (close to) degenerate, the molecule becomes charge bistable[27,64,22]. Considering a phonon bath, the Franck-Condon overlap gives rise to the (nearly) Gaussian-shaped transition probabilities (grey) for tunnelling into excited vibrational states[65]. Tip-sample tunnelling events correspond to transitions that go downward in the many-body diagram (dark blue arrows). The tunnelling rate for the three different $V_G$ is thus proportional to the purple shaded area in the Gaussians. **b**, **d**, **f**, Single-particle pictures corresponding to the many-body diagrams in **a**, **c**, **e**, respectively. Tunnelling is possible to empty or singly occupied orbitals that lie below the Fermi level of the tip. Note that the Gaussians are used to indicate the transition probabilities, they do not indicate the occupancy of the levels, which is for all shown cases the same. The SOMO of CuPc remains in these cases singly occupied because of the large charging energy associated to changing the occupancy of the strongly spatially localized SOMO. Note that the tunnelling barrier also changes slightly depending on $V_G$.

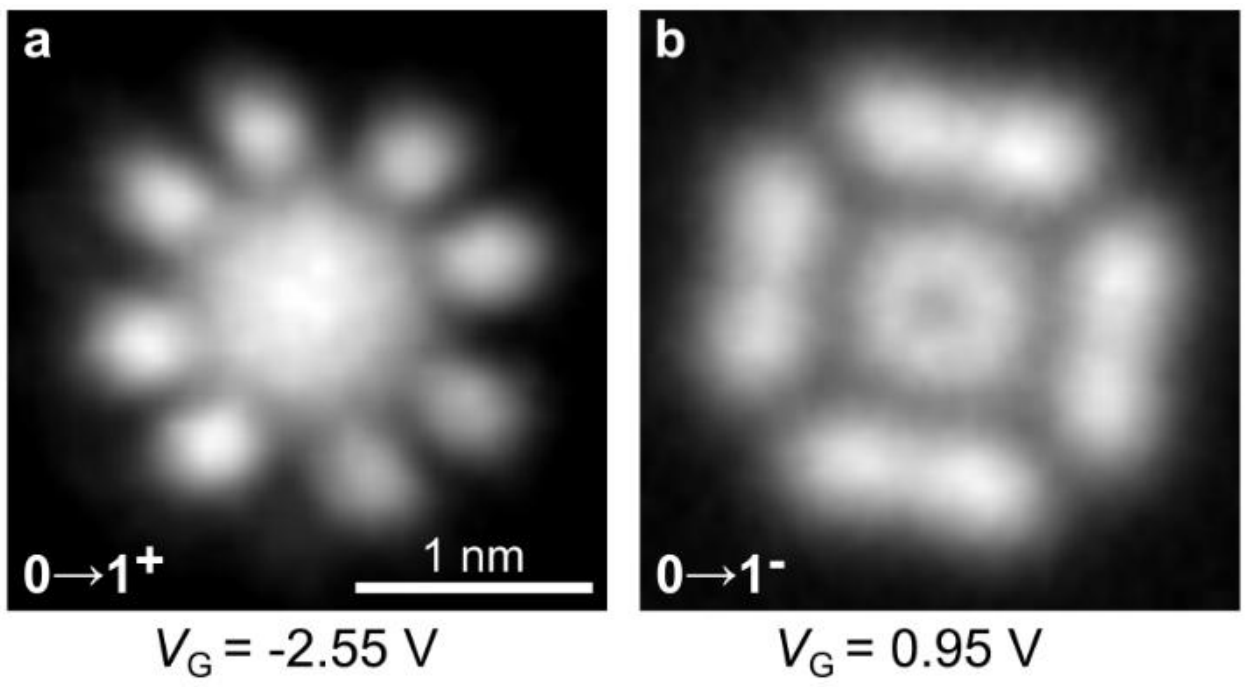

**Extended Data Fig. 4 | AC-STM images of CuPc. a**, **b**, AC-STM images corresponding to the 0→1$^+$ and 0→1$^-$ transitions, respectively. AFM parameters: **a**, $\Delta z$ = -4 Å, $V_G$ = -2.55 V, $V_{a.c.}$ = 1.5 $V_{pp}$, and **b**, $\Delta z$ = -2.4 Å, $V_G$ = 0.95 V, $V_{a.c.}$ = 1.5 $V_{pp}$). $A$ = 1 Å, $\Delta z$ is given with respect to the setpoint $\Delta f$ = -5.5 Hz at $V$ = -2.6 V (**a**) and $V$ = 0.95 V (**b**).

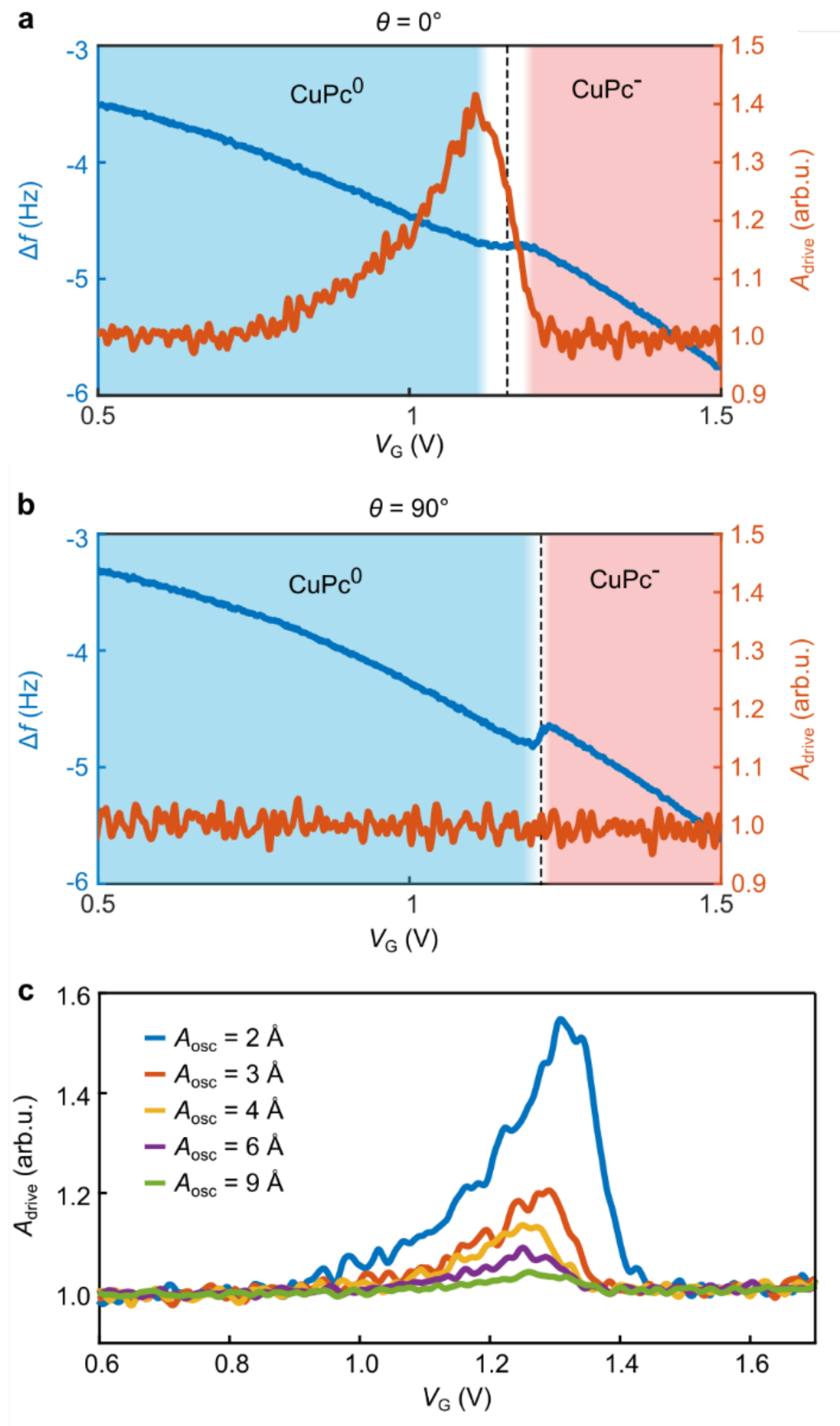

**Extended Data Fig. 5 | Photocurrent signal located at the charge-state transition and its cantilever oscillation amplitude dependence**. **a**, **b**, PE-AFM $A_{\mathrm{drive}}$ signal and simultaneously measured $\Delta f$ as a function of $V_G$. The step in $\Delta f$ (indicated by the dotted line) indicates a change in charge state. **c**, PE-AFM $A_{\mathrm{drive}}$ signal as a function of $V_G$ for different cantilever oscillation amplitudes, as indicated. These measurements were performed for a CuPc molecule adsorbed with its metal centre above a $Cl^-$ site of NaCl(>20 ML)/Cu(111). The tip-height was adjusted such that the closest turnaround point of the tip remained at the same distance to the sample, while the zero crossing moved away from the sample by up to 7 Å. Experimental parameters: CW laser power 8.5 mW. $A$ = 2 Å, $\Delta z$ = -5 Å; $A$ = 3 Å, $\Delta z$ = -4 Å; $A$ = 4 Å, $\Delta z$ = -3 Å; $A$ = 5 Å, $\Delta z$ = -2 Å; $A$ = 6 Å, $\Delta z$ = -1 Å; $A$ = 9 Å, $\Delta z$ = 2 Å. $\Delta z$ is given with respect to the setpoint $\Delta f$ = 1.25 Hz at $V$ = 0 V, $A$ = 1 Å.

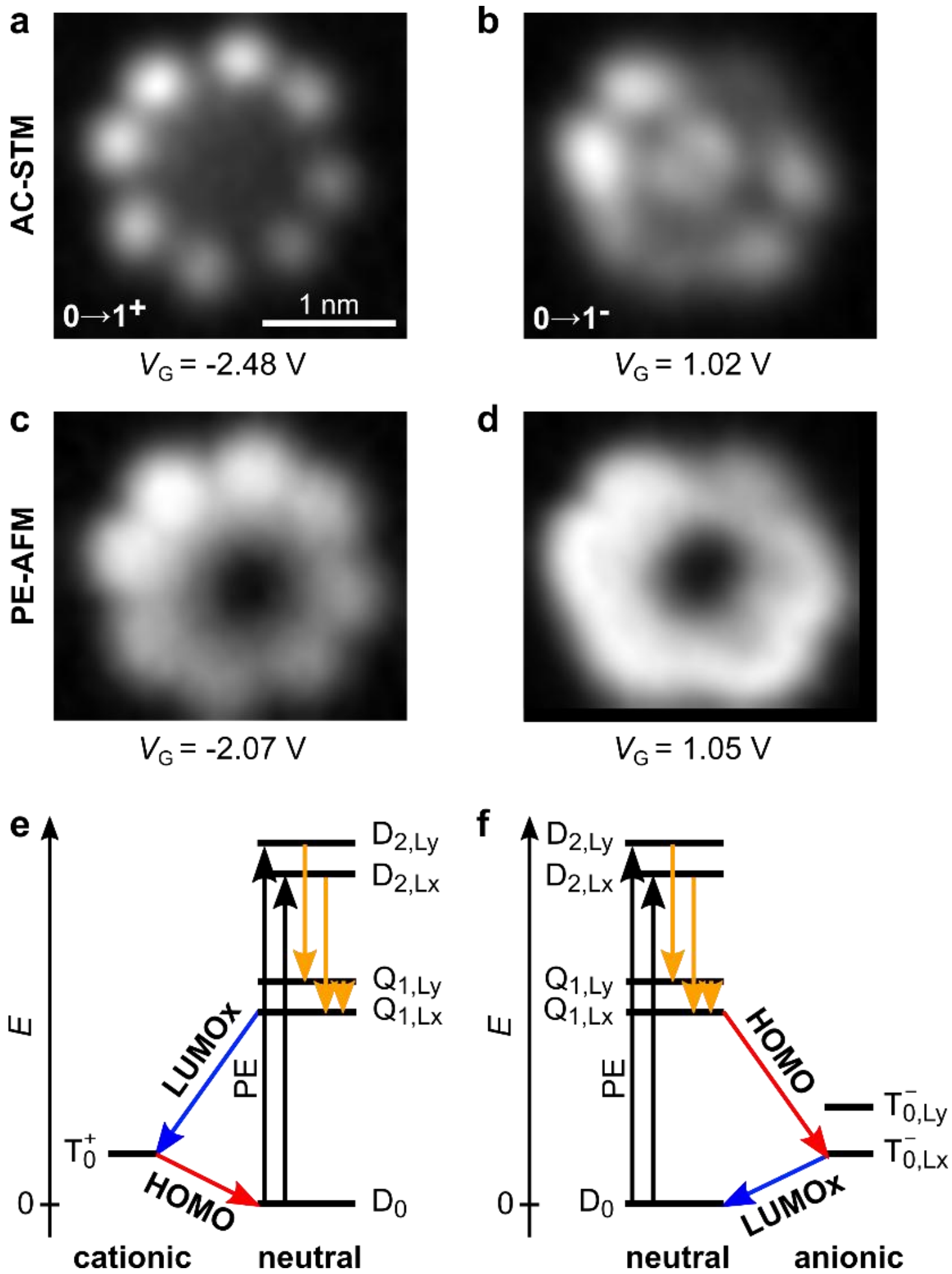

**Extended Data Fig. 6 | Orbital-selective contrast in symmetry-broken CuPc. a**, **b**, AC-STM images corresponding to the 0→1⁺ and 0→1⁻ transition, respectively, for a molecule with lifted LUMO degeneracy in the ground state. **c**, **d**, PE-AFM images on the same molecule taken at negative and positive $V_G$, respectively, reveal orbital-selective contrast. **e**, **f**, Simplified many-body diagram illustrating the relevant states for the photoexcitation contrast generation of panels **c**, **d**, respectively. Experimental parameters: **a**, $A$ = 1 Å, $\Delta z$ = -1.8 Å, $V_G$ = -2.475 V, $V_{a.c.}$ = 1.4 $V_{pp}$, and **b**, $A$ = 1 Å, $\Delta z$ = -2.6 Å, $V_G$ = 1.02 V, $V_{a.c.}$ = 1.4 $V_{pp}$. In case of **a** and **b**, $\Delta z$ is given with respect to the setpoint $\Delta f$ = 1.5 Hz at $V$ = 0 V, $A$ = 1 Å. **c**, $A$ = 1 Å, $\Delta z$ = -1.5 Å, $V_G$ = -2.07 V, CW laser power: 4.8 mW and **d**, $A$ = 1 Å, $\Delta z$ = -4.7 Å, $V_G$ = 1.05 V, CW laser power: 4.8 mW. In case of **c** and **d**, $\Delta z$ is given with respect to the setpoint $\Delta f$ = 1.2 Hz at $V$ = 0 V, $A$ = 1 Å.

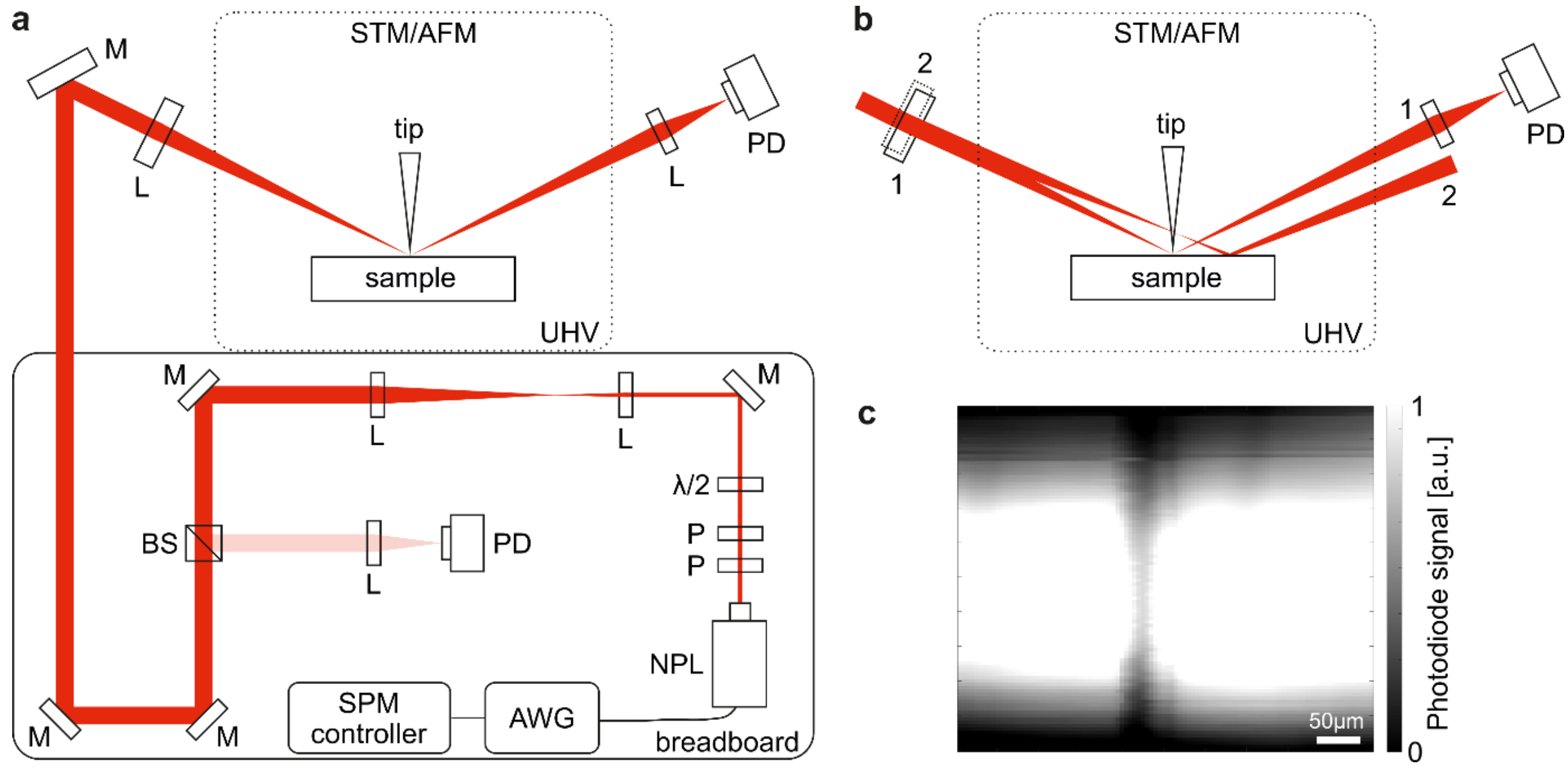

**Extended Data Fig. 7 | Schematic overview of the laser setup and laser alignment. a**, Sketch of the laser setup, which consists of two parts. One part is fixed on a breadboard, which is mounted to the frame of the microscope. The other part is rigidly connected to the UHV chamber of the system. Legend: AWG: arbitrary waveform generator; NPL: nanosecond pulsed laser; P: polarizer; $\lambda/2$: $\lambda/2$ wave plate; M: mirror; L: lens; BS: beam splitter; PD: photodiode. **b**, Sketch illustrating the laser alignment, with two different positions of the focusing lens: Upon proper alignment (lens position 1), the laser is hitting the tip–sample junction. **c**, Typical photodiode signal measured for different positions of the focusing lens. In the middle, the shadow of the tip and its reflection are visible. The scale bar indicates a 50 μm displacement of the focusing lens.

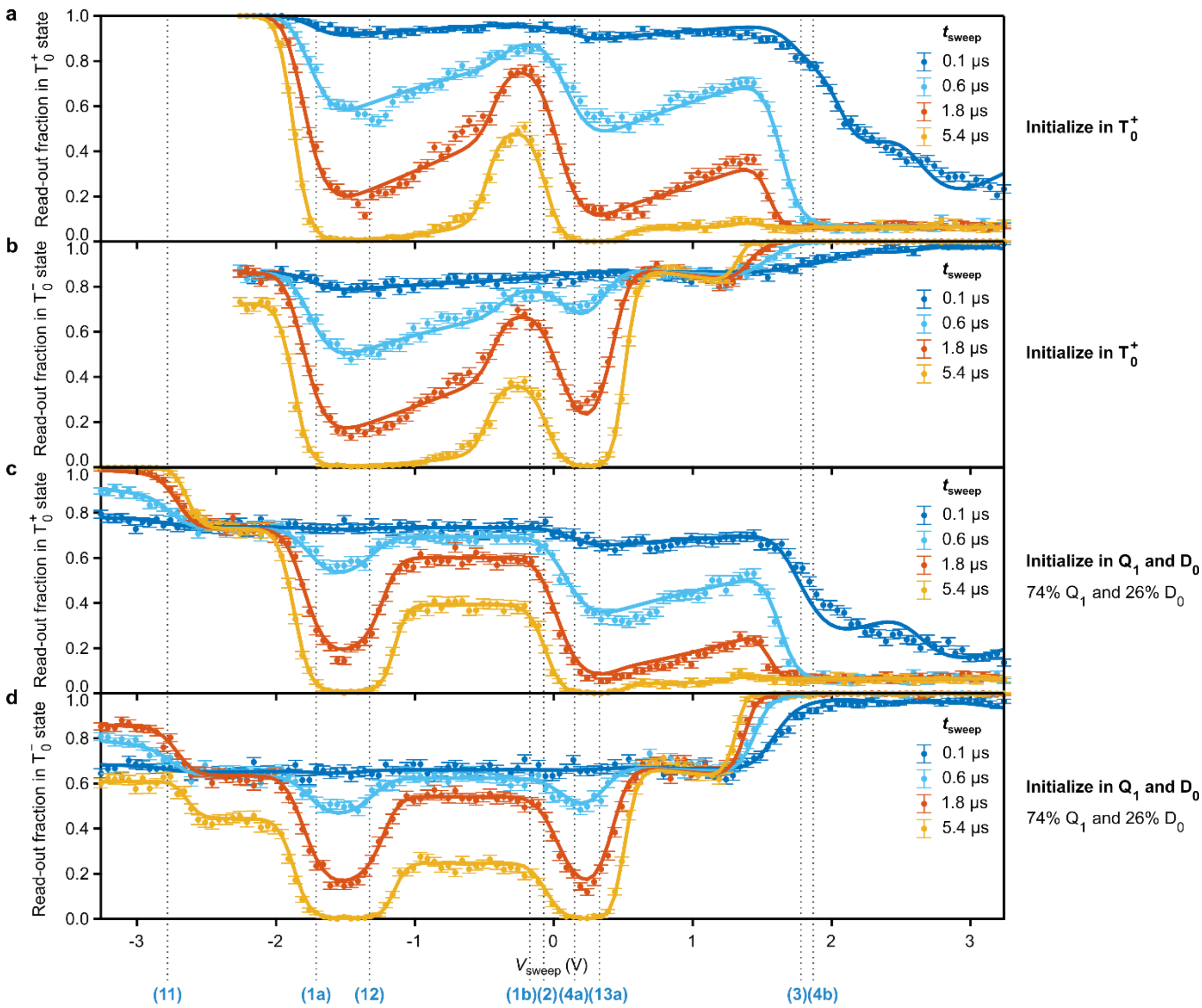

**Extended Data Fig. 8 | Excited-state spectroscopy of CuPc.** Excited-state spectroscopy data measured using a pump-probe voltage pulse sequence consisting of a set, sweep and read-out phase during which specific gate voltages were applied. The set pulse (sequence) initializes the CuPc molecule in the cationic charge state (**a**, **b**, $V_{set}$ = -3.26 V, $t_{set}$ = 333.5 μs) or via the cationic charge state in the neutral states $D_0$ and $Q_1$ (**c**, **d**, $V_{set,1}$ = -3.26 V, $t_{set,1}$ = 333.5 μs and $V_{set,2}$ = -0.26 V, $t_{set,2}$ = 1.8 μs). The duration of the second set pulse was chosen long enough to ensure that the voltage pulse was efficient, but short compared to the lifetime of $Q_1$. Depending on the applied voltage during the subsequent sweep pulse, various electronic states can be occupied. The population of the states at the end of the sweep pulse were mapped onto two different charge states at the beginning of the read-out period, with $V_{read\text{-}out}$ corresponding to the $T_0^+$-$D_0$ degeneracy (**a**, **c**, $V_{read\text{-}out}$ = -2.26 V) or $D_0$-$T_0^-$ degeneracy (**b**, **d**, $V_{read\text{-}out}$ = 0.96 V). A read-out time of 125 ms was used to allow extracting the charge state from the measured AFM frequency shift. The pump-probe voltage pulse sequence was repeated 640 times for each applied sweep voltage to determine the read-out fraction (the normalized population during the read-out phase) in the $T_0^+$ and $T_0^-$ states, respectively. The error bars were derived from the standard deviation for a binominal distribution[29]. Four different $t_{sweep}$ were used, as indicated. Solid lines represent fits to the data. More details can be found in ref. 31, deviations for CuPc are detailed in the Methods section.

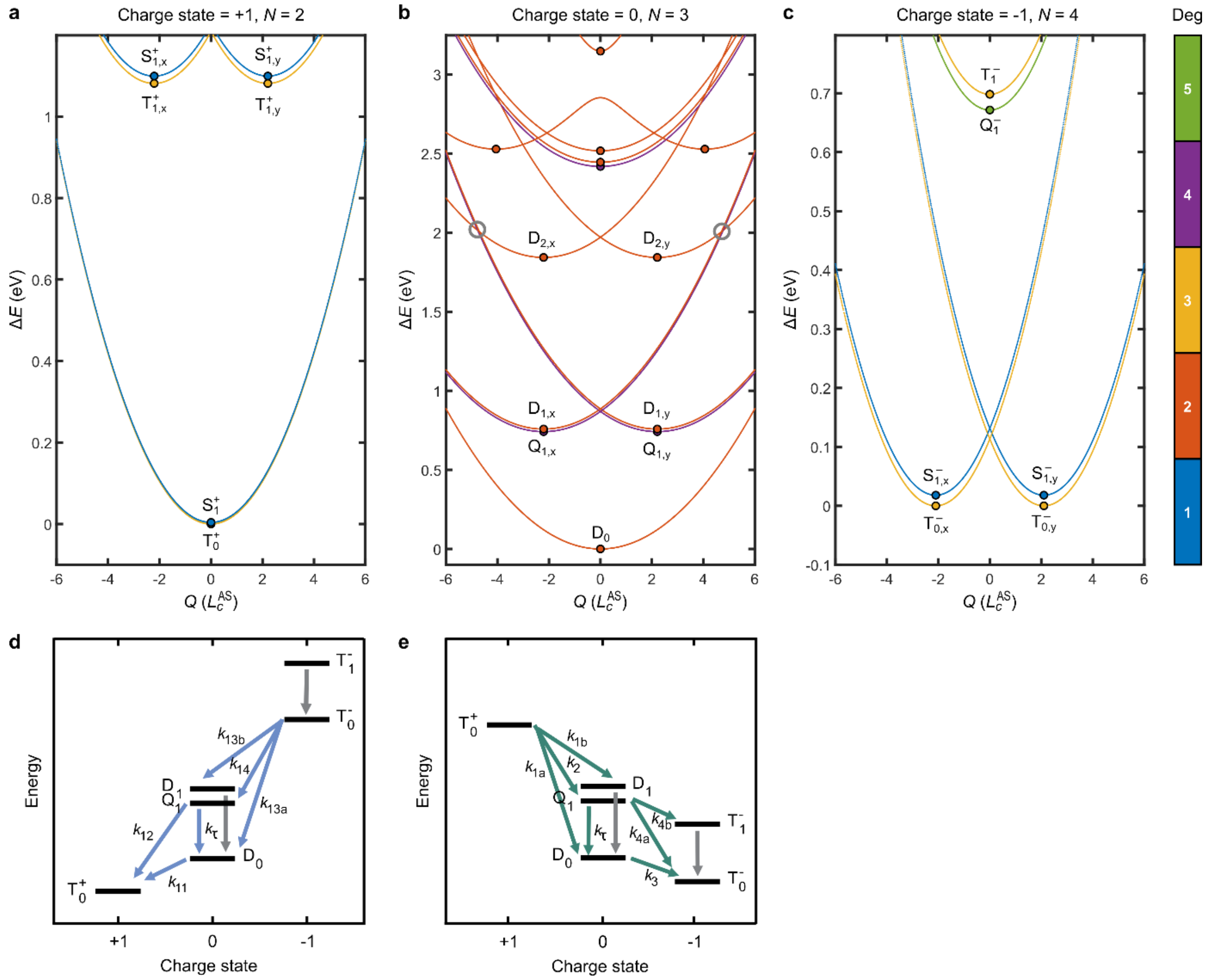

**Extended Data Fig. 9 | Adiabatic potential surfaces for the many-body states of CuPc and possible transitions between them. a-c,** Adiabatic potential surfaces for the many-body states of the cationic, neutral and anionic CuPc, calculated with respectively 2,3, and 4 electrons occupying the frontier orbitals. The horizontal axis represents an effective deformation coordinate along the Jahn-Teller active mode for the combined molecule and substrate (see ref. 62 for reference). A relatively high internal conversion rate of $D_2$ to $D_1$ is expected from the crossings at low excess energies of the APES of $D_{2,x}$ and $D_{1,y}$ as well as the APES of $D_{2,y}$ and $D_{1,x}$, which are indicated by grey circles. **d**, **e**, Many-body diagrams for two applied gate voltages showing the transitions, which are opened at the respective voltage. The gray arrows indicate spin-allowed charge-conserving transitions, which are considered as immediate on the timescale of our experiment. The arrow labelled with $k_\tau$ indicates the decay of $Q_1$ into $D_0$, the other labelled arrows indicate tunnelling transitions with transition rates $k_i$.

| | CuPc 1 centre | CuPc 1 lobe | CuPc 2 centre | Average |
|---|---|---|---|---|
| **Transition voltage (V)** | | | | |
| $T_0^+ \rightarrow D_0$ | -1.71 ± 0.05 | -1.79 ± 0.05 | -1.77 ± 0.05 | -1.76 ± 0.08 |
| $T_0^+ \rightarrow Q_1$ | -0.07 ± 0.10 | -0.12 ± 0.10 | -0.23 ± 0.16 | -0.14 ± 0.16 |
| $T_0^+ \rightarrow D_1$ | -0.17 ± 0.10 | -0.15 ± 0.07 | -0.28 ± 0.15 | -0.20 ± 0.13 |
| $Q_1 \rightarrow T_0^-$ | 0.15 ± 0.05 | 0.08 ± 0.05 | 0.09 ± 0.05 | 0.11 ± 0.07 |
| $Q_1 \rightarrow T_1^-$ | 1.87 ± 0.10 | 1.81 ± 0.10 | 1.81 ± 0.10 | 1.83 ± 0.06 |
| $T_0^- \rightarrow D_0$ | 0.33 ± 0.10 | 0.43 ± 0.10 | 0.32 ± 0.10 | 0.36 ± 0.12 |
| $D_0 \rightarrow T_0^-$ | 1.78 ± 0.05 | 1.71 ± 0.05 | 1.72 ± 0.05 | 1.73 ± 0.08 |
| $Q_1 \rightarrow T_0^+$ | -1.33 ± 0.05 | -1.25 ± 0.05 | -1.36 ± 0.05 | -1.31 ± 0.11 |
| $D_0 \rightarrow T_0^+$ | -2.78 ± 0.05 | -2.69 ± 0.05 | -2.81 ± 0.05 | -2.76 ± 0.13 |
| **Other fitting parameters** | | | | |
| $T_1$ decay rate ($\mu s^{-1}$) | 0.118 ± 0.003 | 0.130 ± 0.004 | 0.122 ± 0.004 | 0.12 ± 0.01 |
| $f_{-\rightarrow+}$ | 0.065 ± 0.003 | 0.090 ± 0.003 | 0.056 ± 0.004 | 0.07 ± 0.03 |
| $f_{+\rightarrow-}$ | 0.967 ± 0.004 | 0.929 ± 0.005 | 0.973 ± 0.005 | 0.96 ± 0.05 |
| Initial $T_1$ population | 0.743 ± 0.003 | 0.734 ± 0.003 | 0.785 ± 0.003 | 0.75 ± 0.05 |
| **Energy difference (eV)** | | | | |
| $Q_1$-$D_0$ | 1.15 ± 0.11 | 1.17 ± 0.11 | 1.08 ± 0.14 | 1.13 ± 0.12 |
| $D_1$-$D_0$ | 1.08 ± 0.11 | 1.15 ± 0.10 | 1.05 ± 0.13 | 1.09 ± 0.13 |
| $T_1^-$-$T_0^-$ | 1.20 ± 0.12 | 1.21 ± 0.12 | 1.20 ± 0.12 | 1.21 ± 0.09 |
| **Reorganization energy (eV)** | | | | |
| $T_0^+$-$D_0$ | 0.75 ± 0.07 | 0.63 ± 0.07 | 0.73 ± 0.07 | 0.70 ± 0.14 |
| $D_0$-$T_0^-$ | 1.02 ± 0.11 | 0.90 ± 0.10 | 0.98 ± 0.10 | 0.96 ± 0.14 |
| $T_0^+$-$Q_1$ | 0.88 ± 0.10 | 0.79 ± 0.10 | 0.79 ± 0.13 | 0.82 ± 0.12 |

**Extended Data Table 1. Fitting results for CuPc.** The results of the fitting for three datasets measured for two CuPc molecules are listed. The error bars on the transition voltages and other fitting parameters are given by the uncertainty on the fit, with a minimal error bar on the transition voltages set to 0.05 V or 0.10 V, as explained in the Methods section. The energy differences and reorganization energies were determined from the fitted voltages. The uncertainties were determined as described in the Methods section. Note that for some states there is quasi-degeneracy, see Extended Data Fig. 9 a-c, and in some cases both of the quasi-degenerate states could be accessed with a single-electron tunnelling process. We assume, therefore, that the transition from $Q_1$ into $T_1^-$ likely also includes the transition into the quintet $Q_1^-$, and the transitions $D_0 \rightarrow T_0^-$ and $D_0 \rightarrow T_0^+$ likely also include $D_0 \rightarrow S_0^-$ and $D_0 \rightarrow S_0^+$, respectively. Note that, depending on their nature and subsequent decay pathways, these transitions affect the fitting parameters corresponding to rates (not listed) differently. Because such quasi-degenerate states are typically not resolved in the fitting procedure, they are modelled as a single transition.